\documentclass[%
 aip, pof,
 amsmath,amssymb,
preprint,%
floatfix,
]{revtex4-1}

\usepackage{graphicx}
\usepackage{dcolumn}
\usepackage{bm}

\usepackage[utf8]{inputenc}
\usepackage[T1]{fontenc}
\usepackage{mathptmx}

\usepackage[nomentbl]{nomencl}
\setnomtableformat{p{0.08\textwidth}p{0.5\textwidth}sp{0.2\textwidth}@{}l}
\newcommand{\mat}  [1]   { \left[ \begin{array}{ccc} #1 \end{array} \right]}

\graphicspath{{Fig/}}

\renewcommand{\nomgroup}[1]{%
\ifthenelse{\equal{#1}{S}}{\item\textbf{Scalars}}{%
\ifthenelse{\equal{#1}{T}}{\item\textbf{Vectors}}{%
\ifthenelse{\equal{#1}{U}}{\item\textbf{Tensors}}{%
\ifthenelse{\equal{#1}{X}}{\item\textbf{Special}}{%
\ifthenelse{\equal{#1}{Y}}{\item\textbf{Subscripts}}{}
}}}}}
\makenomenclature
\begin{document}

\title{A charged finitely extensible dumbbell model: \\ Explaining rheology of dilute polyelectrolyte solutions}

\author{D. Shogin}
 \altaffiliation[Also at ]{The National IOR Centre of Norway, University of Stavanger, 4036 Ullandhaug, Norway} 
 \email{dmitry.shogin@uis.no}
\affiliation{Department of Energy Resources, University of Stavanger, 4036 Ullandhaug, Norway
}%

\author{P. A. Amundsen}
 \altaffiliation[Also at ]{The National IOR Centre of Norway, University of Stavanger, 4036 Ullandhaug, Norway} 
\affiliation{%
Department of Mathematics and Physics, University of Stavanger, 4036 Ullandhaug, Norway
}%

\date{\today}

\begin{abstract}
A robust non-Newtonian fluid model of dilute polyelectrolyte solutions is derived from kinetic theory arguments. Polyelectrolyte molecules are modeled as finitely elongated nonlinear elastic dumbbells, where effective charges (interacting through a simple Coulomb force) are added to the beads in order to model the repulsion between the charged sections of polyelectrolyte chains. It is shown that the relative strength of this repulsion is regulated by the electric-to-elastic energy ratio, $E$, which is one of the key parameters of the model. In particular, $E$ accounts for the intrinsic rigidity of polyelectrolyte molecules and can be used to explain the impact of solvent salinity on polyelectrolyte rheology. 
With two preaveraging approximations, the constitutive equations of the resulting fluid model are formulated in closed form. Material functions predicted by the model for steady shear flow, steady extensional flow, small-amplitude oscillatory shear flow, and start-up and cessation of steady shear flow are obtained and  investigated using a combination of analytical and numerical methods. In particular, it is shown how these material functions depend on $E$. The two limiting cases of the model -- uncharged dumbbells ($E=0$) and rigid dumbbells ($E\to \infty$) -- are  included in the analysis. It is found that despite its simplicity, the  model predicts most of experimentally observed rheological features of polyelectrolyte solutions.
\end{abstract}

\maketitle

\printnomenclature

\section{Introduction}
\label{Sec:Introduction}
Polymeric liquids -- liquids containing extremely long molecules -- are of great importance for modern engineering and biotechnological applications. In contrast to liquids consisting of small molecules, polymeric liquids are non-Newtonian: they do not obey the laws of classical fluid mechanics and often behave strictly opposite to what one could expect from "ordinary" fluids. \cite{Bird1987a,Spagnolie2014}
\par 
Many synthetic and biological polymers are polyelectrolytes: when dissolved in polar liquids (commonly water), they form polyions -- parts of the molecular chain become electrically charged. The charges are of identical sign, which leads to repulsive electric forces acting between different parts of the chain. Such forces make the molecules more rigid and rod-like and therefore mechanically different from their electrically neutral analogs, as described by de Gennes \textit{et al}. \cite{deGennes1976} 
\par 
The physics and chemistry of polyelectrolytes have been a hot research topic during the last few decades, with much attention paid to experiments, theory, and numerical simulations (for details, see the recent review by Muthukumar\cite{Muthukumar2017} and references therein). At the same time, most of the theories consider only equilibrium properties of polyelectrolyte soltuions; very little progress has been made in understanding and modeling their rheology, which is quite specific. One of the crucial features of polyelectrolytes is that their rheological properties are strongly sensitive to the kind and amount of ions present in the solvent. When the charged parts of a polyion attract counterions -- e.g., from a dissolved salt -- the repulsive force acting between these parts decreases due to screening; hence, the polyions become more flexible. \cite{Dou2008,Afolabi2019b} This, of course, strongly affects the rheological behavior of the solution.
\par 
Two major fields where flows of polyelectrolyte solutions are of  importance are biological physics \cite{Spagnolie2014,Visakh2014,Maitz2015, DelGiudice2016} and enhanced oil recovery.  \cite{Lake1989, Wever2011,Jimenez2018, Afolabi2019b,Gbadamosi2019} In both cases, transient flows in complex geometries are involved. Such flows typically have both shear and extensional components, which for non-Newtonian fluids are coupled in a nontrivial way. In addition, the salinity of the solvent may vary, making the flows of polyelectrolytes more complicated compared to those of uncharged polymers. To understand the dynamics of such flows -- even qualitatively -- it is necessary to use advanced tensor models based on microscopic physics\cite{Bird1987b} to adequately describe the forces governing the fluid motion; these models must take the polyelectrolyte nature of the polymer into account. At the same time, such models need to be  simple enough in order to be practically applicable at industrial scales and satisfy the demands of applied technology.
\par 
We should also remark that any fluid model designed to work for complex flows must correctly describe the response of the fluid to simple shear and shearfree flows in both steady and transient regimes and predict the well-known rheological properties of polymer solutions, such as shear-thinning, extensional thickening, and existence of stress overshoots at flow start-up.\cite{Bird1987a}
\par
Dumbbell models \cite{Bird1987b} are a relatively simple class of fluid models for polymer solutions, derived from microscopic kinetic theory. Each polymer molecule is thought of as composed of two identical spherical beads connected by a spring (elastic dumbbell) or by simply a rigid rod (rigid dumbbell). The configuration of each dumbbell is completely specified by the connector vector, $\boldsymbol{Q}$, pointing from one bead to the other. Then, the properties of the molecules are defined by the connector force law, $\boldsymbol{F}_\mathrm{c}(\boldsymbol{Q})$. In particular, the Warner force law \cite{Warner1972} has been found physically relevant,
\begin{equation}
\label{Eq:Model:WarnerLaw}
\boldsymbol{F}_\mathrm{c}(\boldsymbol{Q})=\dfrac{H \boldsymbol{Q}}{1-(Q/Q_0)^2}.
\end{equation}%
\nomenclature[tQ]{$\boldsymbol{Q}$}{connector vector}{m}{first used in Eq. (\ref{Eq:Model:WarnerLaw})}%
\nomenclature[tF]{$\boldsymbol{F}_\mathrm{c}$}{connector force}{N}{first used in Eq. (\ref{Eq:Model:WarnerLaw})}
\nomenclature[sQ2]{$Q$}{dumbbell extension}{m}{first used in Eq. (\ref{Eq:Model:WarnerLaw})}%
\nomenclature[sQ3]{$Q_0$}{dumbbell extension limit}{m}{first used in Eq. (\ref{Eq:Model:WarnerLaw})}%
\nomenclature[sH]{$H$}{Warner spring coefficient}{\mathrm{N}\cdot m^{-1}}{first used in Eq. (\ref{Eq:Model:WarnerLaw})}%
This law describes a nonlinear spring of finite extensibility: the extension of the spring cannot exceed $Q_0$, and the spring is Hookean with stiffness $H$ at small extensions.   Finite extensibility proves to be a crucial property for understanding the flow of polymer solutions. The resulting macroscopic fluid model is known as the FENE (finitely elongated nonlinear elastic) dumbbell model. \cite{Armstrong1974a, Armstrong1974b} A preaveraging approximation (closure) made by Peterlin\cite{Peterlin1966} allows to formulate the constitutive equations in closed form, facilitating both analytical investigation and numerical simulations of the fluid model. This updated version of the model is now known as FENE-P ("P" for Peterlin) and is probably the best dumbbell model to date. It is, however, restricted to electrically neutral or weakly hydrolyzed polymers.
\par 
The pioneering attempt of constructing a dumbbell model describing polyelectrolyte solutions is that of King and Eisenberg,\cite{King1972} who considered a Hookean dumbbell model modified by the presence of effective charges, interacting through an electrostatic Coulomb force. Then, Dunlap and Leal\cite{Dunlap1984} constructed an analogous fluid model based on FENE dumbbells; however, they adopted conformation-dependent friction. No closed-form constitutive equation was derived, and numerical simulations revealed a hysteretic behavior of viscosity and relative extension with respect to flow strength. Ait-Kadi \textit{et al.}\cite{AitKadi1988} formulated a constitutive equation for the model of Dunlap and Leal\cite{Dunlap1984} using a conformation tensor approach. Since then, the focus has been kept on implementing advanced numerical methods and improving the modeling of the electric repulsion between the charged parts of  polyelectrolyte molecules.\cite{Andrews1998,Jiang2001,Zhou2006} \par 
In this work, we are aiming at developing an effective phenomenological fluid model, useful for understanding the rheology of polyelectrolyte solutions and for qualitative description of their behavior in complex flows. Such a model must encapsulate all the crucial features of polyelectrolytes: orientability, nonlinearity, finite extensibility, and variable intrinsic rigidity of the molecules, and at the same time be as simple as possible without suffering from pathologies of early kinetic theory models.  For these reasons, we take the successful FENE-P dumbbell model as our starting point. Although screened Coulombic interactions (in the Debye-H\"{u}ckel approximation) are a proper way to describe the repulsion between charged parts of polyelectrolyte molecules, this would also lead to substantial mathematical complexity. Therefore, we adopt the concept of effective charges, reverting to a simple Coulomb force, as proposed by Dunlap and Leal. \cite{Dunlap1984}
\par 
The paper is organized as follows: In Sec. \ref{Sec:ConstEq}, the underlying assumptions of the new polymer fluid model are specified and the closed-form constitutive equations are derived. In Secs. \ref{Sec:SteadyShearFlow}-\ref{Sec:StartupCessation}, we investigate the resulting fluid model. The constitutive equations are used to obtain the material functions predicted by the model, and the properties of these functions are studied. This is done for several standard steady and transient flows: steady shear flow (Sec. \ref{Sec:SteadyShearFlow}), steady extensional flow (Sec. \ref{Sec:SteadyExtensionalFlow}), small-amplitude oscillatory shear flow (Sec. \ref{Sec:SAOS}), and start-up and cessation of steady shear flow (Sec. \ref{Sec:StartupCessation}). Each of Secs. \ref{Sec:SteadyShearFlow}-\ref{Sec:StartupCessation} is divided into analysis and discussion parts. Finally, conclusions are presented in Sec. \ref{Sec:Conclusions}.
\par
Throughout this paper, SI units are used. Scalars, such as temperature $T$ and shear rate $\dot{\gamma}$, are written with lightface italic font; vectors, such as velocity $\boldsymbol{v}$, with boldface Latin; while second-order tensors, such as rate-of-strain tensor $\boldsymbol{\dot{\gamma}}$ and stress tensor $\boldsymbol{\tau}$, with boldface Greek. For the stress tensor, the sign convention of Bird \textit{et al.} \cite{Bird1987a} is adopted.
\section{The constitutive equations}
\label{Sec:ConstEq}
The FENE dumbbell model is assumed applicable for sufficiently dilute polymer solutions so that the polymer molecules interact with the molecules of the Newtonian solvent much more strongly than with each other. The polymer-solvent interactions are typically described by an isotropic Stokes's law with conformation-independent coefficient $\zeta$. \cite{Bird1987b} It has been found that, macroscopically, the polymer contribution to the stress tensor of the solution depends on three parameters: the "ideal-gas" pressure ($nkT$, where $n$ is the number concentration of  dumbbells, $k$ is Boltzmann's constant, and $T$ is the thermodynamic temperature), the dimensionless nonlinearity parameter,
\begin{equation}
b=HQ^2_0/kT,
\label{Eq:Model:b}
\end{equation}
and a time constant, which we shall find it most convenient to define by
\begin{equation}
\lambda = \lambda_Q= \dfrac{\zeta Q_0^2}{12kT}.
\label{Eq:Model:LambdaQ}
\end{equation}%
\nomenclature[sb]{$b$}{nonlinearity parameter}{-}{defined by Eq. (\ref{Eq:Model:b})}%
\nomenclature[sZyl0]{$\lambda$}{time constant of C-FENE-P dumbbells}{s}{defined by Eq. (\ref{Eq:Model:LambdaQ})}%
\nomenclature[sZyf]{$\zeta$}{hydrodynamic drag coefficient}{\mathrm{kg\cdot s^{-1}}}{first used in Eq. (\ref{Eq:Model:LambdaQ})}%
\nomenclature[sn]{$n$}{number concentration of dumbbells}{\mathrm{m^{-3}}}{first used in Eq. (\ref{Eq:ConstEq:Kramers})}%
\nomenclature[sk]{$k$}{Boltzmann's constant}{\mathrm{J\cdot K^{-1}}}{first used in Eq. (\ref{Eq:Model:b})}%
\nomenclature[sT]{$T$}{thermodynamic temperature}{K}{first used in Eq. (\ref{Eq:Model:b})}%
\nomenclature[sZyl0h]{$\lambda_H$}{time constant of FENE dumbbells}{s}{defined by Eq. (\ref{Eq:Model:LambdaH})}%
\nomenclature[sZyl0q]{$\lambda_Q$}{time constant of rigid dumbbells}{s}{defined by Eq. (\ref{Eq:Model:LambdaQ})}%
More commonly, 
\begin{equation}
\lambda_H =\dfrac{3\lambda_Q}{b}=\dfrac{\zeta}{4H}
\label{Eq:Model:LambdaH}
\end{equation}
is used in the literature. Our alternative choice of the time constant will be discussed below.
\par 
In order to describe qualitatively the electric repulsion between the charged parts of the polyelectrolyte chain, we assume the beads to carry identical effective charges $q$, interacting via an  electrostatic Coulomb force. To describe this repulsive interaction, an extra term is added to the FENE-P connector force so that
\begin{equation}
\label{Eq:Model:ConnectorForce}
\boldsymbol{F}_\mathrm{c} = \dfrac{H \boldsymbol{Q}}{1-(Q/Q_0)^2}-\dfrac{q^2}{4\pi \varepsilon_0 \varepsilon}\dfrac{\boldsymbol{Q}}{Q^3},
\end{equation}
where $\varepsilon_0$ is the permittivity of vacuum and $\varepsilon$ is the relative permittivity of the solvent.%
\nomenclature[sQy]{$q$}{effective bead charge}{C}{first used in Eq. (\ref{Eq:Model:ConnectorForce})}%
\nomenclature[sZye]{$\varepsilon$}{relative permittivity of the solvent}{-}{first used in Eq. (\ref{Eq:Model:ConnectorForce})}%
\nomenclature[sZye0]{$\varepsilon_0$}{permittivity of vacuum}{\mathrm{F\cdot m^{-1}}}{first used in Eq. (\ref{Eq:Model:ConnectorForce})}%
\par 
Similar to other models describing dilute polymer solutions, the stress tensor can be written as a sum of independent solvent and polymer contributions. Since the solvent is assumed Newtonian, only the polymer contribution, $\boldsymbol{\tau}$, is of rheological interest. It can be written in two forms -- the Kramers form and the Giesekus form, respectively, \cite{Bird1987b,Kramers1944,Giesekus1962} 
\begin{align}
\boldsymbol{\tau} &= -n \langle \boldsymbol{QF}_\mathrm{c} \rangle+nkT\boldsymbol{\delta}, \label{Eq:ConstEq:Kramers}\\
\boldsymbol{\tau} &= \dfrac{1}{4}n\zeta \langle \boldsymbol{QQ} \rangle_{(1)}. \label{Eq:ConstEq:Giesekus}
\end{align}%
\nomenclature[ut]{$\boldsymbol{\tau}$}{polymer contribution to  stress tensor}{Pa}{first used in Eq. (\ref{Eq:ConstEq:Kramers})}
\nomenclature[ud]{$\boldsymbol{\delta}$}{unit tensor}{-}{first used in Eq. (\ref{Eq:ConstEq:Kramers})}%
\nomenclature[xZz]{$\langle \rangle$}{configuration-space average}{-}{first used in Eq. (\ref{Eq:ConstEq:Kramers})}%
\nomenclature[yZ]{(1)}{upper-convected time (Oldroyd) derivative}{s^{-1}}{first used in Eq. (\ref{Eq:ConstEq:Giesekus})}%
Here, the angular brackets denote the configuration-space average,\cite{Bird1987b} $\boldsymbol{\delta}$ is the unit tensor, while the subscript "$(1)$" stands for the upper-convected time derivative, introduced by Oldroyd\cite{Oldroyd1950} to express the rate of change in tensor properties of a fluid element in a coordinate system deforming with the fluid.
\par 
Substituting the modified connector force [Eq. (\ref{Eq:Model:ConnectorForce})] into the Kramers expression for the stress tensor [Eq. (\ref{Eq:ConstEq:Kramers})], one gets
\begin{equation}
\boldsymbol{\tau} = -nH \left\langle \dfrac{\boldsymbol{QQ}}{1-(Q/Q_0)^2}\right\rangle+\dfrac{nq^2}{4\pi \varepsilon_0 \varepsilon}\left\langle\dfrac{\boldsymbol{QQ}}{Q^3} \right\rangle+nkT\boldsymbol{\delta}. 
\label{Eq:ConstEq:TauNotPreaveraged}
\end{equation}
The configuration distribution function of the dumbbells, needed to calculate the averages in this equation, cannot be obtained directly. Therefore, in order to arrive at a closed-form constitutive equation, the averages must be eliminated from Eq. (\ref{Eq:ConstEq:TauNotPreaveraged}). This is made possible by preaveraging the first two terms on the right-hand side of the equation, which is implemented as follows:
\begin{align}
\left\langle \dfrac{\boldsymbol{QQ}}{1-(Q/Q_0)^2}\right\rangle & \approx \dfrac{\langle \boldsymbol{QQ} \rangle }{1-\langle Q^2\rangle /Q^2_0}, \label{Eq:ConstEq:ClosureP}\\
\left\langle\dfrac{\boldsymbol{QQ}}{Q^3} \right\rangle & \approx \dfrac{\langle\boldsymbol{QQ}\rangle}{{\langle Q^2 \rangle}^{3/2}}.\label{Eq:ConstEq:ClosureE}
\end{align}
Approximation (\ref{Eq:ConstEq:ClosureP}) is Peterlin's closure, introduced when formulating the FENE-P dumbbell polymer model. \cite{Peterlin1966} The corresponding approximation for the Coulomb term [Eq. (\ref{Eq:ConstEq:ClosureE})] is established by analogous arguments. We propose the abbreviation C-FENE-P for this extended polymer model, where "C" stands for "charged".
\par 
To describe the strength of the electric interactions, it is convenient to introduce the dimensionless ratio,
\begin{equation}
\label{Eq:ConstEq:DefinitionE}
E=\frac{q^2}{(4\pi \varepsilon_0 \varepsilon Q_0) kT},
\end{equation} %
\nomenclature[sE2]{$E$}{electric-to-elastic energy ratio}{-}{defined by Eq. (\ref{Eq:ConstEq:DefinitionE})}%
between the characteristic potential energy of the electric repulsion and the thermal energy scale of the dumbbells. Alternatively, one can write
\begin{equation}
E=z^2 l_\mathrm{B}/Q_0,
\label{Eq:ConstEq:DefinitionEAlt}
\end{equation}
where $l_\mathrm{B}$ is the Bjerrum length\cite{Dou2008} and $z=q/e$ the valence. Larger values of $E$ correspond to increased electrostatic repulsion between the beads, i.e., to "stiffer" dumbbells. In the limit $E\to 0$, the original (uncharged) FENE-P model is recovered, while the dumbbells become rigid as $E \to \infty$.%
\nomenclature[slb]{$l_\mathrm{B}$}{Bjerrum length}{m}{first used in Eq. (\ref{Eq:ConstEq:DefinitionEAlt})}%
\nomenclature[se1]{$e$}{elementary electric charge}{C}{first used in Eq. (\ref{Eq:ConstEq:DefinitionEAlt})}%
\nomenclature[szx]{$z$}{valence}{-}{first used in Eq. (\ref{Eq:ConstEq:DefinitionEAlt})}%
\par 
We further introduce the mean-square relative dumbbell extension, $x$, as
\begin{equation}
\label{Eq:ConstEq:Definitionx}
x=\dfrac{ \langle Q^{2} \rangle }{Q_{0}^{2}},
\end{equation}%
\nomenclature[sx]{$x$}{mean-square relative dumbbell extension}{-}{defined by Eq. (\ref{Eq:ConstEq:Definitionx})}%
which is a measure of dumbbell stretching: $x$ reaches its minimal (non-zero) value at equilibrium and $x \to 1$ at maximal stretching, i.e., when $Q\to Q_0$.
\par 
Making use of the preaveraging approximations   (\ref{Eq:ConstEq:ClosureP}) and (\ref{Eq:ConstEq:ClosureE}) and notations (\ref{Eq:ConstEq:DefinitionE})-(\ref{Eq:ConstEq:Definitionx}), Eq. (\ref{Eq:ConstEq:TauNotPreaveraged}) can be rewritten as
\begin{equation}
\boldsymbol{\tau} = - nH \dfrac{\langle \boldsymbol{QQ} \rangle}{1-x}+nH\dfrac{E}{b}\dfrac{\langle \boldsymbol{QQ} \rangle}{x^{3/2}}+nkT\boldsymbol{\delta}.
\label{Eq:ConstEq:TauPreaveraged}
\end{equation}
Having defined the C-FENE-P $Z$-factor by
\begin{equation}
\label{Eq:ConstEq:DefinitionZ}
Z=\dfrac{1}{1-x}-\dfrac{E}{b}\dfrac{1}{x^{3/2}},
\end{equation} %
\nomenclature[sZ]{$Z$}{$Z$-factor}{-}{defined by Eq. (\ref{Eq:ConstEq:DefinitionZ})}%
one writes Eq. (\ref{Eq:ConstEq:TauPreaveraged}) as
\begin{equation}
\boldsymbol{\tau} =-nHZ\langle \boldsymbol{QQ} \rangle+nkT\boldsymbol{\delta}.
\label{Eq:ConstEq:TauPreaveragedUltimate}
\end{equation}
\par 
This is identical in form to the corresponding equation of the FENE-P dumbbell model. \cite{Bird1980} One can still proceed by taking the Oldroyd derivatives of both sides of the equation, eliminating $\langle \boldsymbol{QQ} \rangle_{(1)}$ using the Giesekus form of the stress tensor (\ref{Eq:ConstEq:Giesekus}) and then eliminating $\langle \boldsymbol{QQ} \rangle$ using Eq. (\ref{Eq:ConstEq:TauPreaveragedUltimate}) once more. The result is
\begin{equation}
\dfrac{b}{3} Z \boldsymbol{\tau} + \lambda \boldsymbol{\tau}_{(1)}-\lambda \left\{ \boldsymbol{\tau}-nkT\boldsymbol{\delta} \right\} \mathrm{D}_t \ln Z=-nkT\lambda \boldsymbol{\dot{\gamma}}, 
\label{Eq:ConstEq:MainConstEq}
\end{equation} %
\nomenclature[st]{$t$}{time variable}{s}{first used in Eq. (\ref{Eq:ConstEq:MainConstEq})}%
\nomenclature[uc]{$\boldsymbol{\dot{\gamma}}$}{rate-of-strain tensor}{\mathrm{s^{-1}}}{defined by Eq. (\ref{Eq:ConstEq:DefinitionRateOfStrain})}%
\nomenclature[tv]{$\boldsymbol{v}$}{fluid velocity}{\mathrm{m\cdot s^{-1}}}{first used in Eq. (\ref{Eq:ConstEq:MainConstEq})}%
where
\begin{equation}
\boldsymbol{\dot{\gamma}} = (\boldsymbol{\nabla v})+(\boldsymbol{\nabla v})^\mathrm{T} 
\label{Eq:ConstEq:DefinitionRateOfStrain}
\end{equation}
is the rate-of-strain tensor, $\boldsymbol{v}$ is the velocity field of the fluid, $t$ is the time variable, while $\mathrm{D}_t$ stands for the material derivative. This result is identical to the constitutive equation of the FENE-P dumbbells. Thus, the difference between the FENE-P and the C-FENE-P models is exclusively the appearance of $E$ in the $Z$-factor.
\par 
Taking the trace of Eq. (\ref{Eq:ConstEq:TauPreaveraged}) and making some simple rearrangements, one arrives at
\begin{equation}
Zx = \dfrac{3}{b}\left( 1-\dfrac{\mathrm{tr}\, \boldsymbol{\tau}}{3nkT} \right).
\label{Eq:ConstEq:TraceEq}
\end{equation}
Combining this with the definition of the $Z$-factor [Eq. (\ref{Eq:ConstEq:DefinitionZ})] leads to the following algebraic equation for $x$:
\begin{equation}
\label{Eq:ConstEq:EquationForx}
\dfrac{1}{1-x}-\frac{E}{b\sqrt{x}}=Z_\mathrm{FENE},
\end{equation}%
\nomenclature[sX]{$X$}{inverse of $x$}{-}{defined by Eq. (\ref{Eq:ConstEq:DefinitionX})}%
where 
\begin{equation}
Z_\mathrm{FENE}=1+\dfrac{3}{b}\left( 1-\dfrac{\mathrm{tr}\, \boldsymbol{\tau}}{3nkT}\right)
\label{Eq:ConstEq:OriginalZ}
\end{equation}
is the $Z$-factor of the original FENE-P dumbbell model.
\par 
In order to facilitate the solution of Eq. (\ref{Eq:ConstEq:EquationForx}), we replace $x$ with its inverse, $X$,
\begin{equation}
\label{Eq:ConstEq:DefinitionX}
X=\dfrac{1}{x}.
\end{equation}
Then,
\begin{equation}
 \dfrac{1}{1-x}=\dfrac{X}{X-1}=1+\dfrac{1}{X-1},
 \end{equation}
and Eq. (\ref{Eq:ConstEq:EquationForx}) becomes
\begin{equation}
(Z_\mathrm{FENE}-1)+\dfrac{E}{b}\sqrt{X}=\dfrac{1}{X-1}.
\label{Eq:ConstEq:EquationForX}
\end{equation}
\par 
We further introduce a function $\mathcal{F}$ of two arguments, $s>0$ and $\alpha \geq 0$, as the unique real solution of the equation
\begin{equation}
s+\alpha \sqrt{y}=\dfrac{1}{y-1}
\label{Eq:ConstEq:EquationForF}
\end{equation}%
\nomenclature[yf]{FENE}{"of FENE-P dumbbells"}{-}{first used in Eq. (\ref{Eq:ConstEq:OriginalZ})}%
\nomenclature[xF]{$\mathcal{F}$}{special function}{-}{defined by Eq. (\ref{Eq:ConstEq:EquationForF})}%
\nomenclature[sZya]{$\alpha$}{dummy argument}{-}{first used in Eq. (\ref{Eq:ConstEq:EquationForF})}%
\nomenclature[sy]{$y$}{dummy variable}{-}{first used in Eq. (\ref{Eq:ConstEq:EquationForF})}%
\nomenclature[ss]{$s$}{dummy argument}{-}{first used in Eq. (\ref{Eq:ConstEq:EquationForF})}%
with respect to $y$. Some properties of this function will be used in the following. In particular, at fixed $s$, $\mathcal{F}$ is  monotonically decreasing with $\alpha$; and at fixed $\alpha$, it is monotonically decreasing as $s$ increases. Furthermore, $\mathcal{F}(s,\alpha)>1$ on its domain, with
\begin{equation}
\lim_{s \to \infty} \mathcal{F}(s,\alpha)=\lim_{\alpha \to \infty} \mathcal{F}(s,\alpha)=1,
\end{equation}
and finally, $\mathcal{F}(s,0) = 1+1/s$. Then, Eq. (\ref{Eq:ConstEq:EquationForX}) can be solved for $X$,
\begin{equation}
X = \mathcal{F}(Z_\mathrm{FENE}-1,E/b),
\end{equation}
while from Eq. (\ref{Eq:ConstEq:TraceEq}), one obtains
\begin{equation}
Z = (Z_\mathrm{FENE}-1)\mathcal{F}(Z_\mathrm{FENE}-1,E/b).
\label{Eq:ConstEq:MainZ}
\end{equation}
The constitutive equations are therefore completely  formulated through expressions (\ref{Eq:ConstEq:MainConstEq}), (\ref{Eq:ConstEq:OriginalZ}), and  (\ref{Eq:ConstEq:MainZ}). 
\par 
One observes that the C-FENE-P dumbbell model contains the following four parameters: $(nkT)$, $b$, $\lambda$, and $E$. The first three are precisely those of the original FENE-P dumbbell model, while $E$ is specific to C-FENE-P and describes the intrinsic rigidity of polyelectrolyte molecules, with larger values of $E$ corresponding to stiffer molecules. It also accounts for the salt-sensitivity of polyelectrolytes, $E$ and the salt concentration in the solvent being inversely related: higher salinity means lower values of $E$ and vice versa.
\par 
Finally, we consider the rigid dumbbell limit, $E\to \infty$. It follows from Eq.  (\ref{Eq:ConstEq:EquationForF}) that $y \to 1$ when $\alpha \to \infty$ for a finite $s$. Thus, $\mathcal{F}(Z_\mathrm{FENE}-1,\infty)=1$ for finite $Z_\mathrm{FENE}$. This is in agreement with physical expectations: an infinitely strong electric repulsion will extend the spring to the upper limit so that $Q\to Q_0$ and $x \to 1$. This leads to a polymer fluid model with the following constitutive equations: 
\begin{align}
& Z_\mathrm{RDB} \boldsymbol{\tau} + \lambda \boldsymbol{\tau}_{(1)}-\lambda \left\{ \boldsymbol{\tau}-nkT\boldsymbol{\delta} \right\} \mathrm{D}_t {\ln Z_\mathrm{RDB}}=-nkT\lambda \boldsymbol{\dot{\gamma}}, 
\label{Eq:ConstEq:RDBConstEq} \\ 
& Z_\mathrm{RDB}=1-\dfrac{\mathrm{tr}\, \boldsymbol{\tau}}{3nkT}.
\label{Eq:ConstEq:RDBZ}
\end{align}%
\nomenclature[yr]{RDB}{"of rigid dumbbells"}{-}{first used in Eq. (\ref{Eq:ConstEq:RDBConstEq})}%
We shall refer to this as the rigid dumbbell (RDB) polymer model. Another rigid dumbbell model, derived using physical assumptions different from ours, was introduced and investigated earlier by Bird \textit{et al.}\cite{Bird1971} The two models share a lot of similarities, but they are not equivalent. A detailed comparison between them lies beyond the scope of this work and shall be discussed elsewhere. In what follows, "RDB" refers to the model formulated by Eqs. (\ref{Eq:ConstEq:RDBConstEq}) and (\ref{Eq:ConstEq:RDBZ}). 
\par 
Note that neither $H$ nor $b$ appears explicitly in Eqs.  (\ref{Eq:ConstEq:RDBConstEq}) and (\ref{Eq:ConstEq:RDBZ}). Moreover, $H$, and therefore $b$, is not defined for the rigid dumbbells. As a result, the commonly adopted microscopic time scale $\lambda_H$ is not applicable in the RDB limit, but $\lambda_Q$, defined by Eq. (\ref{Eq:Model:LambdaQ}), is independent of $H$ and hence provides a universal microscopic time scale for the FENE-P, C-FENE-P, and RDB models. This justifies our choice $\lambda=\lambda_Q$.
\par 
In Secs. \ref{Sec:SteadyShearFlow}-\ref{Sec:StartupCessation}, we shall investigate and analyze the material functions predicted by the C-FENE-P dumbbell and RDB fluid models for some standard flow regimes. The contribution of the Newtonian solvent to the material functions is well understood; therefore, only the polymer contribution to the material functions will be discussed.
\section{Steady shear flow}
\label{Sec:SteadyShearFlow}
\subsection{Analysis}
Steady shear flow can be locally described by a fluid velocity field given by
\begin{equation}
\label{Eq:MF:SSF:Velocity}
\boldsymbol{v} = \begin{bmatrix}
v_1(x_2) & 0 & 0 \end{bmatrix},
\end{equation}
at any position $\boldsymbol{x}$, the stress tensor taking the form
\begin{equation}
\label{Eq:MF:SSF:StressTensor}
\boldsymbol{\tau}=\begin{bmatrix}
\tau_{11} & \tau_{12} & 0 \\
\tau_{12} & \tau_{22} & 0 \\
0 & 0 & \tau_{33}
\end{bmatrix},
\end{equation}
with 
\begin{equation}
\label{Eq:MF:SSF:StressTensorOldroyd}
\boldsymbol{\tau}_{(1)}=
-\begin{bmatrix}
2\tau_{12} & \tau_{22} & 0 \\
\tau_{22} & 0 & 0 \\
0 & 0 & 0
\end{bmatrix}
\dot{\gamma}_{\,12}.
\end{equation}%
\nomenclature[tx]{$\boldsymbol{x}$}{position vector}{m}{first used in Eq. (\ref{Eq:MF:SSF:Velocity})}%
\nomenclature[sN1]{$N_1$}{first normal stress difference}{Pa}{defined by Eq. (\ref{Eq:MF:SSF:FNSD})}%
\nomenclature[sN1]{$N_2$}{second normal stress difference}{Pa}{defined by Eq. (\ref{Eq:MF:SSF:SNSD})}%
\nomenclature[sZyV1]{$\Psi_1$}{first normal stress coefficient}{\mathrm{Pa\cdot s^2}}{defined by Eq. (\ref{Eq:MF:SSF:FNSD})}%
\nomenclature[sZyV91]{$\Psi_2$}{second normal stress coefficient}{\mathrm{Pa \cdot s^2}}{defined by Eq. (\ref{Eq:MF:SSF:SNSD})}%
\nomenclature[sZyg]{$\dot{\gamma}$}{shear rate}{\mathrm{s^{-1}}}{first used in Eq. (\ref{Eq:MF:SSF:ShearStress})}%
\nomenclature[sZyha]{$\eta$}{non-Newtonian viscosity}{\mathrm{Pa\cdot s}}{defined by Eq. (\ref{Eq:MF:SSF:ShearStress})}%
The rate-of-strain tensor has only one independent non-zero component $\dot{\gamma}\equiv\dot{\gamma}_{\,12}=\dot{\gamma}_{\,21}$, and the three standard steady shear flow material functions -- non-Newtonian viscosity ($\eta$), first normal stress coefficient ($\Psi_1$), and second normal stress coefficient ($\Psi_2$) -- are defined, respectively, by
\begin{align}
\tau_{12} & = -\eta(\dot{\gamma})\dot{\gamma}, \label{Eq:MF:SSF:ShearStress}\\
N_1=\tau_{11}-\tau_{22} &= -\Psi_1(\dot{\gamma})\dot{\gamma}^{\,2}, \label{Eq:MF:SSF:FNSD} \\
N_2=\tau_{22}-\tau_{33} &= -\Psi_2(\dot{\gamma})\dot{\gamma}^{\,2}, \label{Eq:MF:SSF:SNSD}
\end{align}
where $N_1$ and $N_2$ are the normal stress differences.
\par 
Substituting Eqs. (\ref{Eq:MF:SSF:StressTensor}) and (\ref{Eq:MF:SSF:StressTensorOldroyd}) into the constitutive equation (\ref{Eq:ConstEq:MainConstEq}) yields
\begin{align}
& \dfrac{b}{3}Z\tau_{11} = 2 \lambda \tau_{12} \dot{\gamma}, \label{Eq:MF:SSF:Red1}\\
& \dfrac{b}{3}Z \tau_{12} = - nkT \lambda \dot{\gamma}, \label{Eq:MF:SSF:Red2}\\
& \tau_{22} = \tau_{33} = 0. \label{Eq:MF:SSF:Red3}
\end{align}
It follows from Eqs. (\ref{Eq:MF:SSF:Red1}) and (\ref{Eq:MF:SSF:Red2}) that $\Psi_1$ is directly proportional to the square of the viscosity, the coefficient of proportionality being independent of $E$; as seen from Eq.  (\ref{Eq:MF:SSF:Red3}), $\Psi_2$ vanishes identically,
\begin{align}
\Psi_1(\dot{\gamma}) &= \dfrac{2}{nkT}\eta^2(\dot{\gamma}), \label{Eq:MF:SSF:Psi1Eta2}\\
\Psi_2(\dot{\gamma}) &= 0.
\end{align}
Having eliminated $\tau_{11}$ from Eqs. (\ref{Eq:MF:SSF:Red1}) and (\ref{Eq:MF:SSF:Red2}), one arrives at the following nonlinear algebraic relation between the shear stress and the shear rate:
\begin{equation}
\label{Eq:MF:SSF:ViscosityEqn}
\left(1+\dfrac{2\mathbb{T}_{12}^{\,2}}{3} \right) \mathcal{F}\left(\dfrac{3+2\mathbb{T}_{12}^{\,2}}{b},\dfrac{E}{b} \right) \mathbb{T}_{12} = -\Lambda,
\end{equation}
\nomenclature[sTij]{$\mathbb{T}_{ij}$}{dimensionless stress tensor components}{-}{defined by Eq. (\ref{Eq:MF:SRT:DefinitionTij})}%
where $\mathbb{T}_{12}=\tau_{12}/(nkT)$ is the dimensionless shear stress and $\Lambda = \lambda \dot{\gamma}$ is the dimensionless shear rate. Equation (\ref{Eq:MF:SSF:ViscosityEqn}) can be solved numerically to calculate the steady shear flow properties of the C-FENE-P dumbbells for arbitrary values of $b$ and $E$. 
\par
\begin{figure}
\begin{center}
\includegraphics[width=3.37in]{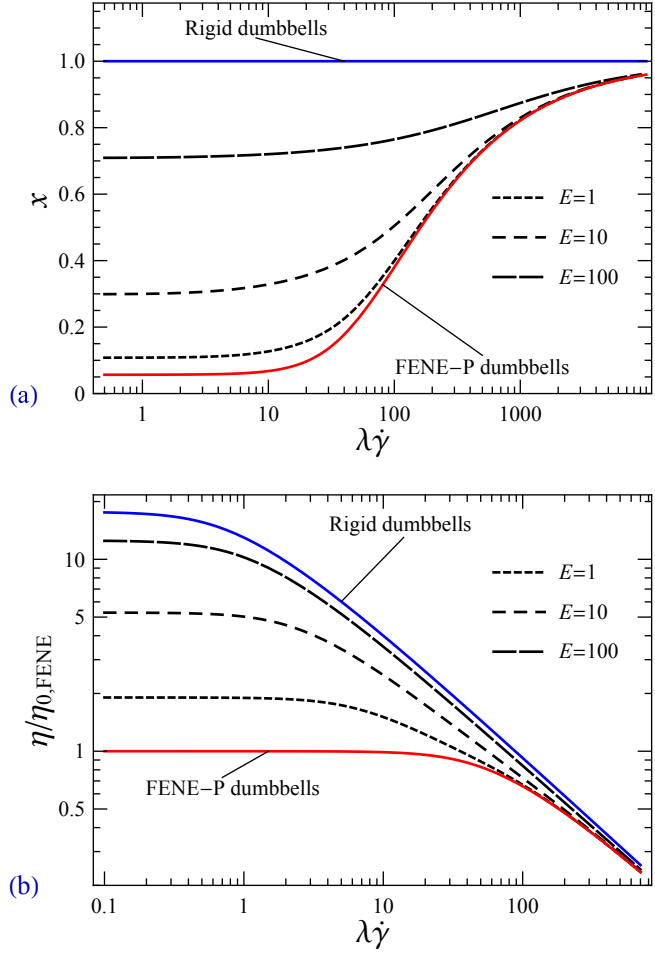} 
\caption{Mean-square relative extension, $x$, in steady shear flow (a) and scaled polymer contribution to non-Newtonian viscosity (b) of C-FENE-P dumbbells, plotted as functions of dimensionless shear rate, $\lambda \dot{\gamma}$, for different values of $E$. The limiting cases $E=0$ (uncharged FENE-P dumbbells) and $E\to \infty$ (rigid dumbbells) are marked. The nonlinearity parameter, $b$, is set to a moderate value of $50$. \label{Fig:SSF-Viscosity}}
\end{center}
\end{figure}

 
The influence of $E$ on the relative extension of the dumbbells in steady shear flow is shown in Fig. \ref{Fig:SSF-Viscosity}(a). In general, a higher value of $E$ leads to larger spring extensions, as expected. This effect is more pronounced at equilibrium and at low-to-medium shear rates. At higher shear rates, the dumbbells are already stretched almost to the upper limit by the flow so that the influence of $E$ becomes small.
\par 
The predicted impact of salinity on the non-Newtonian viscosity is shown in Fig.  \ref{Fig:SSF-Viscosity}(b). The C-FENE-P dumbbells are shear-thinning, which is typical for polymer solutions. The impact of $E$ on viscosity is twofold. First, increasing solvent salinity (decreasing $E$) at any fixed shear rate leads to a reduction in viscosity. This reduction is largest at small shear rates and decreases as the shear rate increases, vanishing as $\dot{\gamma} \to \infty$. Second, the onset of shear-thinning is shifted towards higher shear rates.
\par 
The asymptotic behavior of the viscosity curves governed by Eq. (\ref{Eq:MF:SSF:ViscosityEqn}) can be studied analytically. At very low shear rates, the viscosity approaches its zero-shear-rate value, 
\begin{equation}
\label{Eq:MF:SSF:ZSRViscosity}
\eta_0 = \dfrac{nkT\lambda}{\mathcal{F}(3/b,E/b)},
\end{equation}
which is highly sensitive to $E$. In the FENE-P limit ($E=0$), this result simplifies to
\begin{equation}
\eta_{0,\mathrm{FENE}}=\dfrac{3}{b+3}nkT\lambda,
\end{equation}
and in the RDB limit ($E \to \infty$), the zero-shear-rate viscosity is
\begin{equation}
\eta_{0,\mathrm{RDB}}=nkT \lambda.
\end{equation}%
\nomenclature[sZyhb]{$\eta_0$}{zero-shear-rate viscosity}{\mathrm{Pa \cdot s}}{first used in Eq. (\ref{Eq:MF:SSF:ZSRViscosity})}%
At very high shear rates, the asymptotic behavior of the viscosity curves of the RDB, C-FENE-P (independent of $E$), and FENE-P models is the same,
\begin{equation}
\eta \approx \sqrt[3]{\dfrac{3}{2}}nkT\lambda \Lambda^{-2/3}.
\end{equation}
\par 
Finally, in the FENE-P and RDB limits, the exact analytical solutions of Eq. (\ref{Eq:MF:SSF:ViscosityEqn}) can be obtained. The exact analytical solution for the FENE-P dumbbells is given, e.g., by Shogin \textit{et al.}\cite{Shogin2017a} In the RDB limit, Eq. (\ref{Eq:MF:SSF:ViscosityEqn}) reduces to the following cubic equation for $\mathbb{T}_{12}$:
\begin{equation}
\label{Eq:MF:SSF:ViscosityEqnRDB}
\left(1+\dfrac{2\mathbb{T}_{12}^{\,2}}{3} \right) \mathbb{T}_{12} = -\Lambda.
\end{equation}
This equation has one real solution, which can be obtained, e.g., using Cardano's method, as described by La Nave and Mazur.\cite{LaNave2002} The result can be written as
\begin{equation}
\dfrac{\eta_\mathrm{RDB}}{\eta_{0,\mathrm{RDB}}} = \dfrac
{-2^{1/3} + \left(3 \Lambda+\sqrt{2+9\Lambda^2} \right)^{2/3}}
{2^{2/3} \Lambda \left(3\Lambda+\sqrt{2+9\Lambda ^2} \right)^{1/3}}.
\end{equation}
\nomenclature[sZyLa]{$\Lambda$}{dimensionless shear rate (Sec. \ref{Sec:SteadyShearFlow} only)}{-}{first used in Eq. (\ref{Eq:MF:SSF:ViscosityEqn})}%
\nomenclature[sZyLb]{$\Lambda$}{dimensionless elongation rate (Sec. \ref{Sec:SteadyExtensionalFlow} only)}{-}{defined by Eq. (\ref{Eq:MF:SEF:DefinitionOfLambda})}%
\nomenclature[sZyLc]{$\Lambda$}{dimensionless step-rate value (Sec. \ref{Sec:StartupCessation} only)}{-}{defined by Eq. (\ref{Eq:MF:SRT:DefinitionLambda})}%
\subsection{Discussion}
The influence of salinity on the non-Newtonian viscosity of polyelectrolytes has been subject to extensive experimental investigations. Both a decrease in viscosity and a shift of the onset of shear-thinning towards higher shear rates with increasing salt concentration are well-known features of polyelectrolyte solutions. \cite{AitKadi1987,Tam1989,Tam1990, Vink1992, Dou2008,Wyatt2011,
Stavland2013, Xiong2014, Stanislavskiy2018, Walter2019} It is also reported that polyelectrolytes containing more intrinsically rigid molecules demonstrate a larger shear-thinning. The C-FENE-P dumbbell model predicts these trends, as shown in Fig. \ref{Fig:SSF-Viscosity}(b).
\par 
The asymptotic value of the shear-thinning exponent ($-2/3$) is identical to that of the FENE-P dumbbell and FENE-P bead-spring-chain models\cite{Bird1980} and matches experimental data for dilute partially hydrolyzed polyacrylamides. \cite{Lozhkina2018} One should note, however, that the slope of the log-log viscosity curve of the C-FENE-P dumbbells [see Fig. \ref{Fig:SSF-Viscosity}(b)] is changing gradually with the shear rate ranging from $-2/3$ to $0$. The C-FENE-P model can therefore explain the shear-thinning exponent values in this range.
\par 
Much less data are available on the normal stress coefficients and their dependence on the salt concentration. The shape of the $\Psi_1(\dot{\gamma})$ curve predicted by the C-FENE-P model is realistic and matches the qualitative description given by Bird \textit{et al.}\cite{Bird1987a} The simple nonlinear relation between $\Psi_1(\dot{\gamma})$ and $\eta(\dot{\gamma})$ [Eq.  (\ref{Eq:MF:SSF:Psi1Eta2})] has been tested experimentally for partially hydrolyzed polyacrylamides in a recent study by Lozhkina\cite{Lozhkina2018}; this relation was proven to be qualitatively correct for solutions of high-molecular-weight polyacrylamides but to not hold for their lower-molecular-weight counterparts.
\par 
The second normal stress coefficient vanishes in many kinetic theory-based polymer fluid models,\citep{Bird1980} including the C-FENE-P dumbbells. In practice, it is reported that $|\Psi_2(\dot{\gamma})|<<|\Psi_1(\dot{\gamma})|$; hence, the second normal stress difference does not play a significant role for most flows of practical interest.\cite{Bird1987a}
\section{Steady extensional flow}
\label{Sec:SteadyExtensionalFlow}
\subsection{Analysis}
The steady simple shearfree flow velocity field is
\begin{equation}
\label{Eq:MF:SEF:VelocityField}
\boldsymbol{v}=\begin{bmatrix} -\dfrac{1}{2}\dot{\varepsilon} x_1 &  -\dfrac{1}{2}\dot{\varepsilon} x_2 & \dot{\varepsilon} x_3 \end{bmatrix},
\end{equation}%
\nomenclature[sZye1]{$\dot{\varepsilon}$}{elongation rate}{\mathrm{s^{-1}}}{first used in Eq. (\ref{Eq:MF:SEF:VelocityField})}%
where $\dot{\varepsilon}$ is the time-independent elongation rate, which can take positive and negative values. Equation (\ref{Eq:MF:SEF:VelocityField}) defines uniaxial extension at $\dot{\varepsilon}>0$ and biaxial stretching at $\dot{\varepsilon}<0$.
\par 
The rate-of-strain tensor, the stress tensor, and the Oldroyd derivative of the latter are all diagonal,
\begin{align}
\boldsymbol{\dot{\gamma}} & = \mathrm{diag} (-\dot{\varepsilon},-\dot{\varepsilon},2\dot{\varepsilon}), \label{Eq:MF:SEF:RateOfStrainTensor}\\
\boldsymbol{\tau} & =\mathrm{diag} (\tau_{11}, \tau_{22}, \tau_{33}), \label{Eq:MF:SEF:StressTensor} \\
\boldsymbol{\tau}_{(1)} & = \mathrm{diag} (\tau_{11}, \tau_{22}, -2\tau_{33} )\dot{\varepsilon}, \label{Eq:MF:SEF:StressTensorOldroyd}
\end{align}
with $\tau_{11}=\tau_{22}$ due to the flow symmetry. The only material function characterizing the fluid in this type of flow is the extensional viscosity. Following Bird \textit{et al.},\cite{Bird1987a} we denote it by $\bar{\eta}$ and define by:
\begin{equation}
\label{Eq:MF:SEF:DefinitionOfExtensionalViscosity}
\tau_{33}-\tau_{11} = -\bar{\eta}(\dot{\varepsilon})\dot{\varepsilon}.
\end{equation}%
\nomenclature[sZyhg]{$\bar{\eta}$}{extensional viscosity}{\mathrm{Pa\cdot s}}{defined by Eq. (\ref{Eq:MF:SEF:DefinitionOfExtensionalViscosity})}%
\nomenclature[sZyhh]{$\bar{\eta}_0$}{zero-elongation-rate extensional viscosity}{\mathrm{Pa\cdot s}}{first used in Eq. (\ref{Eq:MF:SEF:TroutonRelation})}%
\par 
Substituting Eqs. (\ref{Eq:MF:SEF:RateOfStrainTensor})-(\ref{Eq:MF:SEF:StressTensorOldroyd}) into the constitutive equation leads to
\begin{align}
\dfrac{b}{3}Z \tau_{11} + \lambda \dot{\varepsilon} \tau_{11} &= nkT \lambda \dot{\varepsilon}, \label{Eq:MF:SEF:Algebraic1} \\
\dfrac{b}{3}Z \tau_{33}-2\lambda \dot{\varepsilon} \tau_{33} &= - 2nkT\lambda \dot{\varepsilon}. \label{Eq:MF:SEF:Algebraic2}
\end{align}
Having replaced $\tau_{11}$, $\tau_{33}$, and $\dot{\varepsilon}$ with dimensionless quantities
\begin{align}
\mathbb{T} & = \dfrac{2 \tau_{11}+\tau_{33}}{nkT}, \label{Eq:MF:SEF:DefinitionOfT}\\
\mathbb{D} &= \dfrac{\tau_{11}-\tau_{33}}{nkT}, \label{Eq:MF:SEF:DefinitionOfD}\\
\Lambda &= \lambda \dot{\varepsilon}, \label{Eq:MF:SEF:DefinitionOfLambda}
\end{align} %
\nomenclature[sTa]{$\mathbb{T}$}{dimensionless trace of the stress tensor}{-}{defined by Eq. (\ref{Eq:MF:SEF:DefinitionOfT})}%
\nomenclature[xD]{$\mathrm{D}_t$}{material derivative}{s^{-1}}{first used in Eq. (\ref{Eq:ConstEq:MainConstEq})}%
\nomenclature[xd]{$\partial_t$}{ordinary time derivative}{s^{-1}}{first used in Eq. (\ref{Eq:MF:SAOS:StressTensorOldroyd})}%
\nomenclature[xZz]{$\boldsymbol{\nabla}$}{del operator}{m^{-1}}{first used in Eq. (\ref{Eq:ConstEq:DefinitionRateOfStrain})}%
\nomenclature[sDb]{$\mathbb{D}$}{dimensionless normal stress difference}{-}{defined by Eq. (\ref{Eq:MF:SEF:DefinitionOfD})}%
one arrives after simple rearrangements at
\begin{align}
\dfrac{b}{3} Z \mathbb{T} + 2 \Lambda \mathbb{D} &= 0, \label{Eq:MF:SEF:DTFormulationEquation1} \\
\dfrac{b}{3} Z \mathbb{D} + \Lambda (\mathbb{T} - \mathbb{D}) &= 3 \Lambda.
\label{Eq:MF:SEF:DTFormulationEquation2}
\end{align}
\begin{figure}
\begin{center}
\includegraphics[width=3.37in]{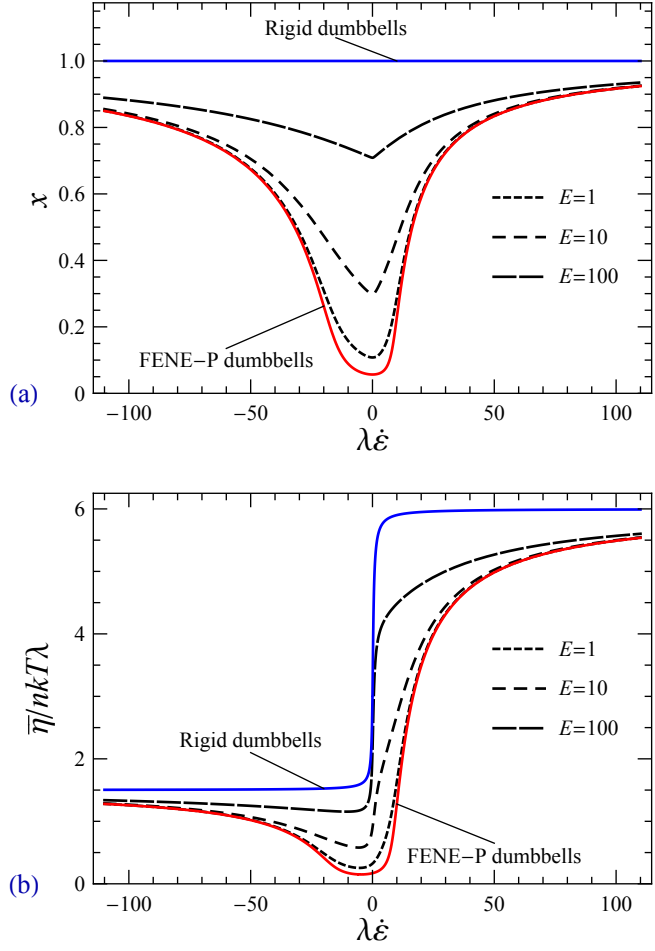} 
\caption{Mean-square relative extension, $x$, in steady extensional flow (a) and scaled polymer contribution to extensional viscosity (b) of C-FENE-P dumbbells, plotted against dimensionless elongation rate, $\lambda \dot{\varepsilon}$, at different values of $E$. The limiting cases $E=0$ (FENE-P dumbbells) and $E\to \infty$ (rigid dumbbells) are marked. Nonlinearity parameter $b=50$.  \label{Fig:SEF-ElongationalViscosity}}
\end{center}
\end{figure} 
\par 
Equations (\ref{Eq:MF:SEF:DTFormulationEquation1}) and (\ref{Eq:MF:SEF:DTFormulationEquation2}) can be solved numerically. The impact of $E$ on the extensional viscosity of the C-FENE-P dumbbells is illustrated in Fig. \ref{Fig:SEF-ElongationalViscosity}(b). It is seen that the C-FENE-P dumbbells show extensional thickening at $\dot{\varepsilon}>0$. An increase in salinity (hence, a decrease in $E$) leads to an overall drop in the extensional viscosity. The local minimum in $\bar{\eta}$ at negative elongation rates, which vanishes in the RDB limit, becomes more pronounced as salinity increases.
\par 
One can also keep track of the mean-square relative extension, $x$, which is shown in Fig. \ref{Fig:SEF-ElongationalViscosity}(a). The extension is smallest at equilibrium and follows the trends similar to those of $\bar{\eta}$ at $\dot{\varepsilon} >0$.
\par 
Some properties of the extensional viscosity curves can be obtained by analytical means. In particular, Eq. (\ref{Eq:MF:SEF:DTFormulationEquation2}) can be used to calculate the zero-elongation-rate extensional viscosity, $\bar{\eta}_0$. Considering the limit $\dot{\varepsilon} \to 0$ and then applying Eqs. (\ref{Eq:MF:SSF:ZSRViscosity}) and (\ref{Eq:MF:SEF:DefinitionOfExtensionalViscosity}) yield
\begin{equation}
\label{Eq:MF:SEF:TroutonRelation}
\bar{\eta}_0=3\eta_0,
\end{equation}
as expected.
\par 
At very large positive or negative elongation rates (in the limit $|\dot{\varepsilon}|\to \infty$), the relative extension of the dumbbells approaches one; hence, the asymptotic behavior of the C-FENE-P and FENE-P dumbbells must be identical to that of the RDB model. In the RDB limit, Eqs.  (\ref{Eq:MF:SEF:DTFormulationEquation1}) and (\ref{Eq:MF:SEF:DTFormulationEquation2}) become
\begin{align}
\left (1 - \dfrac{\mathbb{T}}{3} \right) \mathbb{T} + 2 \Lambda \mathbb{D} &= 0, \\
\left (1 - \dfrac{\mathbb{T}}{3} \right) \mathbb{D} + \Lambda (\mathbb{T}-\mathbb{D}) &= 3\Lambda.
\end{align}
This system has three solutions of which only one,
\begin{align}
\mathbb{T} & = \dfrac{3}{2} \left(1-\Lambda-\sqrt{1-2\Lambda+9\Lambda^2} \right), \\
\mathbb{D} &= \dfrac{3}{4} \left(-1+5\Lambda +\sqrt{1-2\Lambda+9\Lambda^2} \right), \label{Eq:MF:SEF:RDBDTTrueSolution}
\end{align}
provides $\mathbb{T} = \mathbb{D} = 0$ at equilibrium ($\Lambda = 0$) and hence is physically relevant. It is seen that $\mathbb{D} \sim 6 \Lambda $ at $\Lambda \to \infty$, and $\mathbb{D} \sim 3/2 \Lambda$ at $\Lambda \to -\infty$. This means that for all the models under consideration (FENE-P, C-FENE-P, and RDB), $\bar{\eta}$ approaches $\bar{\eta}_{+\infty}=6nkT\lambda$ at large positive elongation rates and $\bar{\eta}_{-\infty}=(3/2)nkT\lambda$ at large negative elongation rates. Equation (\ref{Eq:MF:SEF:RDBDTTrueSolution}) also provides an exact analytical expression for the extensional viscosity in the RDB limit,
\begin{equation}
\dfrac{\bar{\eta}(\Lambda)}{\bar{\eta}_{0,\mathrm{RDB}}}=\dfrac{-1+5\Lambda +\sqrt{1-2\Lambda+9\Lambda^2}}{4\Lambda}. 
\end{equation}
\subsection{Discussion}
The C-FENE-P model predicts monotonic extensional thickening at positive elongation rates. The S-shaped curves, obtained in earlier theoretical works of Dunlap and Leal\cite{Dunlap1984} and Ait-Kadi \textit{et al.},\cite{AitKadi1988} are not reproduced by our model. The impact of salinity on the extensional viscosity, as shown in Fig. \ref{Fig:SEF-ElongationalViscosity}, is in qualitative agreement with experimentally observed trends.\cite{Miles1983, AitKadi1987, Dunlap1987, Ferguson1990, Anna1997,Walter2019} For the negative elongation rates (biaxial stretching), we found no experimental results for comparison.
\section{Small-amplitude oscillatory shear flow}
\label{Sec:SAOS}
\subsection{Analysis}
In small-amplitude oscillatory shear (SAOS) flow, the fluid  velocity field is given by
\begin{equation}
\boldsymbol{v} = 
\begin{bmatrix}
\dot{\gamma}_{\,12}(t) x_2 & 0 & 0 \end{bmatrix},
\label{Eq:MF:SAOS:VelocityField}
\end{equation}
where the harmonically oscillating shear rate
\begin{equation}
\label{Eq:MF:SAOS:ShearRate}
\dot{\gamma}_{\,12}(t)=\dot{\gamma}_{\,21}(t)=\dot{\gamma}_{\,0} \cos(\omega t)
\end{equation}%
\nomenclature[sZyg01]{$\dot{\gamma}_{\,0}$}{shear rate amplitude (Sec. \ref{Sec:SAOS} only)}{\mathrm{s^{-1}}}{first used in Eq. (\ref{Eq:MF:SAOS:ShearRate})}%
\nomenclature[sZyg02]{$\dot{\gamma}_{\,0}$}{step-rate value (Sec. \ref{Sec:StartupCessation} only)}{\mathrm{s^{-1}}}{first used in Eq. (\ref{Eq:MF:SRT:ShearRate})}%
\nomenclature[sZyz]{$\omega$}{angular frequency}{\mathrm{s^{-1}}}{first used in Eq. (\ref{Eq:MF:SAOS:ShearRate})}%
is the only independent non-zero component of the rate-of-strain tensor. Here, $\dot{\gamma}_{\,0}$ is the amplitude of oscillations  and  $\omega$ is the angular frequency. The former is assumed very small so that the dependency of the stress tensor components on $\dot{\gamma}_{\,0}$ is completely described by the lowest-order terms (first order for the shear stresses and second order for the normal stresses). The stress tensor of a fluid undergoing SAOS flow has the general form
\begin{equation}
\label{Eq:MF:SAOS:StressTensor}
\boldsymbol{\tau}(t)=\mat{
\tau_{11}(t) & \tau_{12}(t) & 0 \\
\tau_{12}(t) & \tau_{22}(t) & 0 \\
0 & 0 & \tau_{33}(t)
},
\end{equation}
with 
\begin{equation}
\label{Eq:MF:SAOS:StressTensorOldroyd}
\boldsymbol{\tau}_{(1)}=\partial_t \boldsymbol{\tau}-\mat{
2\tau_{12}(t) & \tau_{22}(t) & 0 \\
\tau_{22}(t) & 0 & 0 \\
0 & 0 & 0
}\dot{\gamma}_{\,12}(t),
\end{equation}
where $\partial_t$ stands for ordinary time derivative.
\par
For polymer solutions, the shear stress oscillates around zero with frequency $\omega$; in contrast to Newtonian liquids, the oscillations are not in phase with those of the shear rate. The normal stress differences oscillate around generally non-zero mean values with the double frequency $2\omega$. \cite{Bird1987a} The properties of the fluid in SAOS flow are then described by eight material functions -- $\eta'$, $\eta''$; $\Psi_1^\mathrm{d}$, $\Psi'_1$, $\Psi''_1$; and $\Psi_2^\mathrm{d}$, $\Psi'_2$, $\Psi''_2$ -- defined by
\begin{align}
\tau_{12} &= -\eta'(\omega) \dot{\gamma}_{\,0} \cos (\omega t)-\eta''(\omega) \dot{\gamma}_{\,0} \sin (\omega t), \label{Eq:MF:SAOS:DefinitionComplexViscosity}\\
N_1 &= - \Psi_1^\mathrm{d}(\omega) \dot{\gamma}_{\,0}^{\,2} - \Psi'_1(\omega) \dot{\gamma}_{\,0}^{\,2} \cos (2\omega t) - \Psi''_1(\omega) \dot{\gamma}_{\,0}^{\,2} \sin(2\omega t), \label{Eq:MF:SAOS:DefinitionComplexFNSC}\\
N_2 &= - \Psi_2^\mathrm{d}(\omega) \dot{\gamma}_{\,0}^{\,2} - \Psi'_2 (\omega)\dot{\gamma}_{\,0}^{\,2} \cos (2\omega t) - \Psi''_2(\omega) \dot{\gamma}_{\,0}^{\,2} \sin(2\omega t).\label{Eq:MF:SAOS:DefinitionComplexSNSC}
\end{align} %
\nomenclature[sZyhc]{$\eta'$}{in-phase component of the complex viscosity}{\mathrm{Pa\cdot s}}{defined by Eq. (\ref{Eq:MF:SAOS:DefinitionComplexViscosity})}%
\nomenclature[sZyhd]{$\eta''$}{out-of-phase component of the complex viscosity}{\mathrm{Pa\cdot s}}{defined by Eq. (\ref{Eq:MF:SAOS:DefinitionComplexViscosity})}%
\nomenclature[sZyV3]{$\Psi_1^\mathrm{d}$}{first normal stress displacement coefficient}{\mathrm{Pa\cdot s^2}}{defined by Eq. (\ref{Eq:MF:SAOS:DefinitionComplexFNSC})}%
\nomenclature[sZyV4]{$\Psi'_1$}{real part of the complex first normal stress \hfill \newline coefficient}{\mathrm{Pa\cdot s^2}}{defined by Eq. (\ref{Eq:MF:SAOS:DefinitionComplexFNSC})}%
\nomenclature[sZyV5]{$\Psi''_1$}{imaginary part of the complex first normal stress \newline coefficient}{\mathrm{Pa\cdot s^2}}{defined by Eq. (\ref{Eq:MF:SAOS:DefinitionComplexFNSC})}%
\nomenclature[sZyV92]{$\Psi_2^\mathrm{d}$}{second normal stress displacement coefficient}{\mathrm{Pa\cdot s^2}}{defined by Eq. (\ref{Eq:MF:SAOS:DefinitionComplexSNSC})}%
\nomenclature[sZyV93]{$\Psi'_2$}{real part of the complex second normal stress \hfill \newline coefficient}{\mathrm{Pa\cdot s^2}}{defined by Eq. (\ref{Eq:MF:SAOS:DefinitionComplexSNSC})}%
\nomenclature[sZyV94]{$\Psi''_2$}{imaginary part of the complex second normal stress \hfill \newline coefficient}{\mathrm{Pa\cdot s^2}}{defined by Eq. (\ref{Eq:MF:SAOS:DefinitionComplexSNSC})}%
The combined quantities $\eta'+\mathrm{i} \eta''$, $\Psi'_1+\mathrm{i} \Psi''_1$, and $\Psi'_2+\mathrm{i} \Psi''_2$ are commonly referred to as complex viscosity, first normal stress coefficient, and second normal stress coefficient, respectively, while $\Psi_1^\mathrm{d}$ and $\Psi_2^\mathrm{d}$ are the normal stress displacement coefficients.
\par 
The in-phase ("real") component of complex viscosity, $\eta'$, describes the direct response of the fluid and can be associated with energy loss due to dissipation. The out-of-phase ("imaginary") component, $\eta''$, arises because the long polymer molecules do not react instantly to rapid flow changes. This leads to a latency, described by a phase shift. This latency can be interpreted as "elasticity" of the flow and associated with energy storage. \cite{Ferry1980,Mezger2014} The complex viscosity components, $\eta'$ and $\eta''$, can also be replaced by the storage and loss moduli, $G'$ and $G''$, defined, respectively, by
\begin{align}
G' &= \eta'' \omega, \label{Eq:MF:SAOS:StorageModulus} \\
G''&=\eta' \omega. \label{Eq:MF:SAOS:LossModulus}
\end{align}%
\nomenclature[sGa]{$G'$}{storage modulus}{Pa}{defined by Eq. (\ref{Eq:MF:SAOS:StorageModulus})}%
\nomenclature[sGb]{$G''$}{loss modulus}{Pa}{defined by Eq. (\ref{Eq:MF:SAOS:LossModulus})}%
These ("elastic") moduli are often measured in experiments.
\par
\begin{figure}
\begin{center}
\includegraphics[width=3.37in]{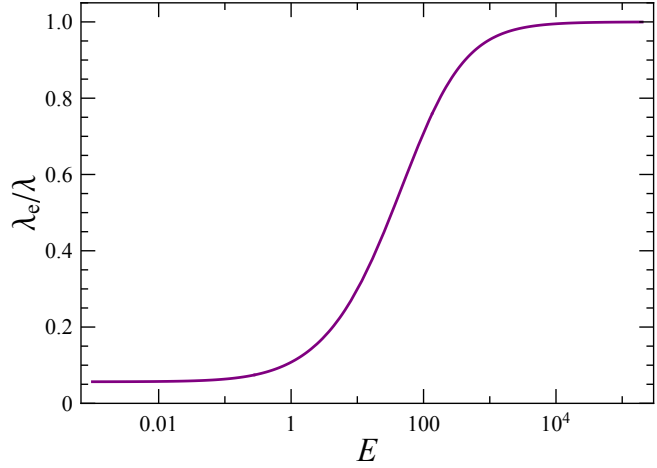} 
\caption{Scaled experimental time constant of C-FENE-P dumbbells, $\lambda_\mathrm{\,e}/\lambda$, as a function of parameter $E$. For illustrative purpose, $b=50$.\label{Fig:SAOS-Lambda}}
\end{center}
\end{figure}

 
The expressions for the SAOS flow material functions of the C-FENE-P dumbbells can be obtained analytically. Substituting Eqs. (\ref{Eq:MF:SAOS:StressTensor}) and (\ref{Eq:MF:SAOS:StressTensorOldroyd}) into the constitutive equation (\ref{Eq:ConstEq:MainConstEq}), keeping only the lowest-order terms in $\dot{\gamma}_{\,12}$, combining and rearranging the scalar equations, one gets
\begin{align}
& N_1 + \lambda_\mathrm{\,e} \partial_t N_1-2 \lambda_\mathrm{\,e} \tau_{12}\dot{\gamma}_{\,12}=0, \label{Eq:MF:SAOS:GeneralFNSD} \\
& N_2+ \lambda_\mathrm{\,e} \partial_t N_2=0, \label{Eq:MF:SAOS:GeneralSNSD}\\
& \tau_{12}+\lambda_\mathrm{\,e} \partial_t \tau_{12} = -\eta_0 \dot{\gamma}_{\,12}, \label{Eq:MF:SAOS:GeneralVisc}
\end{align} %
\nomenclature[sZyl0e]{$\lambda_{\,\mathrm{e}}$}{"experimental" time constant}{s}{defined by Eq. (\ref{Eq:MF:SAOS:LambdaE})}%
where we have introduced the "experimental" time constant $\lambda_\mathrm{\,e}$ by
\begin{equation}
\label{Eq:MF:SAOS:LambdaE}
\lambda_\mathrm{\,e} (b, E) = \dfrac{\eta_0}{nkT}=\dfrac{\lambda}{\mathcal{F}(3/b,E/b)},
\end{equation} 
with $\eta_0$ being the zero-shear-rate viscosity of the C-FENE-P dumbbells given by Eq. (\ref{Eq:MF:SSF:ZSRViscosity}). As shown in Fig. \ref{Fig:SAOS-Lambda},  this time constant is a monotonically increasing function of $E$, ranging from $3\lambda/(b+3)$ at $E=0$ to $\lambda$ at $E\to \infty$. 
\par
\begin{figure}
\begin{center}
\includegraphics[width=8.5cm]{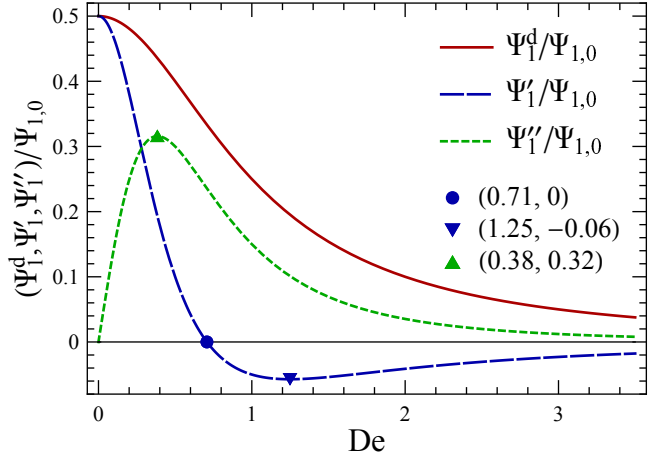} 
\end{center}
\caption{Scaled material functions describing first normal stress difference of C-FENE-P dumbbells in SAOS flow, plotted against the SAOS Deborah number, $\mathrm{De}=\lambda_\mathrm{\,e} \omega$. \label{Fig:SAOS-Scaled}}
\end{figure}

 
Substituting Eqs. (\ref{Eq:MF:SAOS:DefinitionComplexViscosity})-(\ref{Eq:MF:SAOS:DefinitionComplexSNSC}) into Eqs.  (\ref{Eq:MF:SAOS:GeneralFNSD})-(\ref{Eq:MF:SAOS:GeneralVisc}) results in identities, holding for all $(\omega t)$. This leads to eight algebraic equations for the SAOS flow material functions. Then, the latter can be written in scaled form as functions of the oscillarory-flow Deborah number,\cite{Reiner1964,Saengow2015}
\begin{equation}
\mathrm{De}=\lambda_\mathrm{\,e} \omega.
\label{Eq:MF:SAOS:DeborahNumber}
\end{equation}
The linear viscoelastic response is
\begin{align}
\dfrac{\eta'}{\eta_0} &= \dfrac{1}{1+\mathrm{De}^2}, \label{Eq:MF:SAOS:ScaledEtaPrime} \\ 
\dfrac{\eta''}{\eta_0} &= \dfrac{\mathrm{De}}{1+\mathrm{De}^2},  \label{Eq:MF:SAOS:ScaledEtaDPrime} \\
\dfrac{G'}{nkT} &= \dfrac{\mathrm{De}^2}{1+\mathrm{De}^2},\label{Eq:MF:SAOS:ScaledGPrime}  \\
\dfrac{G''}{nkT} & = \dfrac{\mathrm{De}}{1+\mathrm{De}^2},\label{Eq:MF:SAOS:ScaledGDPrime}
\end{align}%
\nomenclature[sDe]{$\mathrm{De}$}{SAOS Deborah number}{-}{defined by Eq. (\ref{Eq:MF:SAOS:DeborahNumber})}%
which is identical to that of a Maxwell fluid with viscosity $nkT\lambda_\mathrm{\,e}$ and time constant $\lambda_\mathrm{\,e}$. \cite{Bird1987a}
The material functions related to $N_2$ -- $\Psi_2^\mathrm{d}$, $\Psi'_2$, and $\Psi''_2$ -- all vanish, while $N_1$ is described by
\begin{align}
\dfrac{\Psi_1^\mathrm{d}}{\Psi_{1,0}} &= \dfrac{1}{2(1+\mathrm{De}^2)},\label{Eq:MF:SAOS:ScaledPsiD}\\
\dfrac{\Psi'_1}{\Psi_{1,0}} &= \dfrac{1-2 \mathrm{De}^2}{2(1+\mathrm{De}^2)(1+4\mathrm{De}^2)}, \label{Eq:MF:SAOS:ScaledPsiPrime}\\
\dfrac{\Psi''_1}{\Psi_{1,0}}&= \dfrac{3\mathrm{De}}{2(1+\mathrm{De}^2)(1+4\mathrm{De}^2)},\label{Eq:MF:SAOS:ScaledPsiDoublePrime}
\end{align} %
\nomenclature[sZyV2]{$\Psi_{1,0}$}{zero-shear-rate first normal stress coefficient}{Pa\cdot s^2}{first used in Eq. (\ref{Eq:MF:SAOS:ScaledPsiD})}%
where $\Psi_{1,0}=2nkT\lambda_\mathrm{\,e}^2$ is the zero-shear-rate first normal stress coefficient of the C-FENE-P dumbbells. 
The scaled material functions, given by Eqs. (\ref{Eq:MF:SAOS:ScaledPsiD})-(\ref{Eq:MF:SAOS:ScaledPsiDoublePrime}), are shown in Fig. \ref{Fig:SAOS-Scaled}, where they are plotted against $\mathrm{De}$.

\par 
The first normal stress displacement coefficient, $\Psi_1^\mathrm{d}$, is a decreasing function of frequency. At low frequencies, $\Psi_1^\mathrm{d}\to \Psi_{1,0}/2$, and at high frequencies, $\Psi_1^\mathrm{d} \sim \omega^{-2}$.
\par  
The real component of the complex first normal stress coefficient, $\Psi'_1$, is the only SAOS material function of the C-FENE-P model, which can take negative values. At very low frequencies, $\Psi'_1\to \Psi_{1,0}/2$. At low-to-moderate frequencies, $\Psi'_1$ decreases with frequency, becoming zero at $\mathrm{De}=1/\sqrt{2}\approx0.707$, and continues to decline until the minimum point,
\begin{equation}
\Psi'_1 = \left(\dfrac{2\sqrt{2}}{3}-1 \right)\Psi_{1,0} \approx -0.057\Psi_{1,0},
\end{equation}
is reached at \begin{equation}
\mathrm{De} = \left[\dfrac{3+\sqrt{2}}{2\sqrt{2}} \right]^{1/2}\approx 1.249.
\end{equation}
Thereafter, $\Psi'_1$ starts to increase with frequency, approaching zero from below, with $\Psi'_1\sim \omega^{-2}$ at high frequencies.
\par 
The imaginary component of the complex first normal stress coefficient, $\Psi''_1$, increases linearly at low frequencies ($\Psi''_1\sim \omega$), reaching a maximum value of 
\begin{equation}
\Psi''_1 =
\dfrac{9\left[3(\sqrt{73}-5)/2 \right]^{1/2} }{23+5\sqrt{73}} \Psi_{1,0} \approx 0.316\Psi_{1,0}
\end{equation}
at 
\begin{equation}
\mathrm{De} = \dfrac{1}{2} \left[ \dfrac{\sqrt{73}-5}{6} \right]^{1/2}\approx 0.384,
\end{equation}
and decays quickly at large frequencies ($\Psi''_1\sim \omega^{-3}$).
\par 
Expressions (\ref{Eq:MF:SAOS:ScaledEtaPrime})-(\ref{Eq:MF:SAOS:ScaledPsiDoublePrime}) show that the SAOS material functions of the C-FENE-P, FENE-P, and RDB fluid models can be written in the same form. However, this form is not suitable for visualizing the impact of $E$, since all the scaling factors depend on $E$. To investigate the $E$-dependence, we reformulate Eqs. (\ref{Eq:MF:SAOS:ScaledEtaPrime})-(\ref{Eq:MF:SAOS:ScaledPsiDoublePrime}) using salinity-independent scaling factors: the zero-shear-rate viscosity and the zero-shear-rate first normal stress coefficient of the FENE-P dumbbells, $\eta_{0,\mathrm{FENE}}$ and $\Psi_{1,0,\mathrm{FENE}}$, in place of $\eta_0$ and $\Psi_{1,0}$, respectively, and $\lambda$ in place of $\lambda_\mathrm{\,e}$.
\par
\begin{figure}
\begin{center}
\includegraphics[width=6.45in]{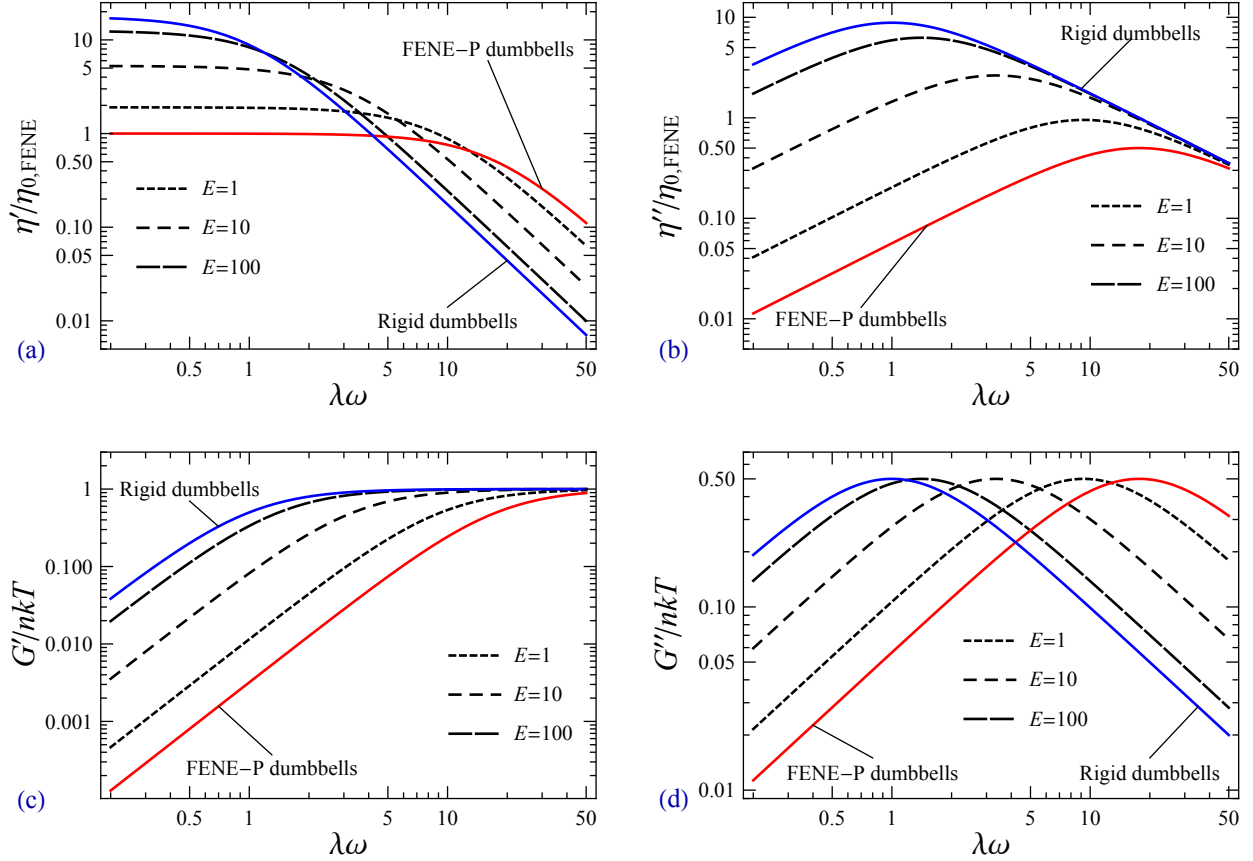}
\caption{Scaled polymer contribution to complex viscosity components [(a) and (b)] and to elastic moduli [(c) and (d)] of C-FENE-P dumbbells, plotted against dimensionless frequency, $\lambda \omega$, for different values of $E$. The limiting cases $E=0$ (FENE-P dumbbells) and $E\to \infty$ (rigid dumbbells) are also shown. Nonlinearity parameter $b=50$. \label{Fig:SAOS-ComplexViscosity}}
\end{center}
\end{figure}

 
The dependence of the in-phase complex viscosity component on $E$ is shown in Fig. \ref{Fig:SAOS-ComplexViscosity}(a). It is seen that a decrease in $E$ leads to a reduction in $\eta'$ at lower frequencies but to an increase in $\eta'$ at higher frequencies; and the onset of "frequency-thinning" is shifted towards higher frequency values as $E$ decreases. 
\par 
The impact of $E$ on the out-of-phase complex viscosity component is visualized in Fig. \ref{Fig:SAOS-ComplexViscosity}(b). A decrease in $E$ mostly affects the low- and mid-frequency regions of the curves: the values of $\eta''$ are reduced, and the maximum is shifted towards higher frequencies.  At very high frequencies, $E$ has no effect on $\eta''$: the curves of the C-FENE-P, FENE-P, and RDB models are asymptotically identical.
\par 
The storage and loss moduli are found to depend on $E$ in a simple way [see Figs. \ref{Fig:SAOS-ComplexViscosity}(c) and \ref{Fig:SAOS-ComplexViscosity}(d)]. If the curves are plotted using a log-log scale, a reduction in $E$ results in a translation of both $G'$ and $G''$ curves to the right due to a decrease in the time constant; the maximum values of both moduli thus remain unchanged.
\begin{figure}
\begin{center}
\includegraphics[width=3.37in]{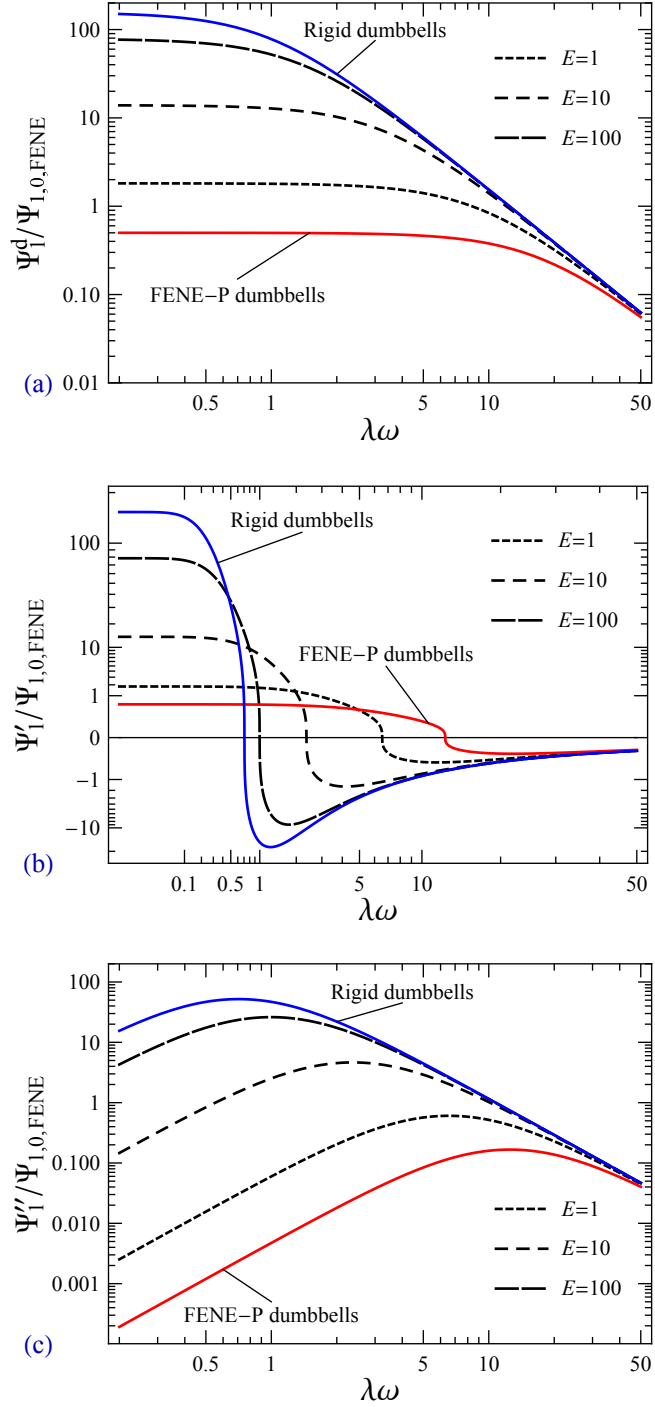} 
\caption{Scaled first normal stress displacement coefficient (a) and complex first normal stress coefficient components [(b) and (c)] of C-FENE-P dumbbells, plotted against dimensionless frequency, $\lambda \omega$, for different values of $E$. The limiting cases $E=0$ (FENE-P dumbbells) and $E\to \infty$ (rigid dumbbells) are also shown. Both axes of plot (b) are scaled with a cubic root function, since $\Psi'_1$ changes its sign. \label{Fig:SAOS-Psi1Primes}}
\end{center}
\end{figure}
\par 
The $E$-dependence of the SAOS material functions related to $N_1$ is shown in Fig. \ref{Fig:SAOS-Psi1Primes}. At low and moderate frequencies, $\Psi_1^\mathrm{d}$ is affected in a way similar to $\eta$ [see Fig. \ref{Fig:SAOS-Psi1Primes}(a)], while $\Psi''_1$ is affected in a way similar to $\eta''$ [see Fig. \ref{Fig:SAOS-Psi1Primes}(c)].
The impact of $E$ on $\Psi'_1$ is more complex but still follows the same general trend: a decrease in $E$ leads to an overall reduction in the magnitude of $\Psi'_1$ and shifts the characteristic points towards higher frequencies [see Fig. \ref{Fig:SAOS-Psi1Primes}(b)]. At very high frequencies, all three material functions are insensitive to $E$.  
\subsection{Discussion}
The qualitative shape and asymptotic behavior of the complex viscosity components $\eta'(\omega)$, $\eta''(\omega)$ and the elastic moduli $G'(\omega)$, $G''(\omega)$, as predicted by the C-FENE-P dumbbell model, are fully consistent with physical arguments provided by Bird \textit{et al.}\cite{Bird1987a} A decrease in these material functions with increasing salinity at fixed $\omega$, as shown in Fig. \ref{Fig:SAOS-ComplexViscosity}, is an experimentally observed feature of polyelectrolyte solutions.\cite{Tam1989,Ihebuzor2019} 
\par 
Furthermore, the C-FENE-P model predicts a decrease of  the experimental time parameter, $\lambda_\mathrm{\,e}$, with increasing salinity (see Fig. \ref{Fig:SAOS-Lambda}); this leads to a shift of the characteristic points, occurring at fixed Deborah numbers, towards higher frequencies. This prediction is compatible with the experimental results of Ihebuzor,\cite{Ihebuzor2019} who reported the $G'$-$G''$ crossover frequency to increase with salinity.
\par 
At the same time, the C-FENE-P model is not capable of resolving the quantitative relations between the SAOS material functions, in particular, it predicts that the $G'(\omega)$ and $G''(\omega)$ curves intersect at the point of maximum of $G''(\omega)$, as follows from Eqs. (\ref{Eq:MF:SAOS:ScaledGPrime}) and  (\ref{Eq:MF:SAOS:ScaledGDPrime}). This is not observed in experiments: as seen from the recent report of Ihebuzor,\cite{Ihebuzor2019} $G''(\omega)$ continues to increase at frequency values larger than the crossover frequency. This mismatch, however, is to be expected: dumbbell models, with their single relaxation time, cannot properly describe the complex oscillatory motion of a real polymer molecule, with a wide spectrum of relaxation times, in cases when the flow pattern changes rapidly. An extension of the C-FENE-P dumbbell model to its bead-spring-chain variant might resolve this issue but would involve significant mathematical and computational complexity, which we are intentionally trying to avoid in this work.
\par 
Finally, we did not find any experimental works allowing for an analysis of the results related to the impact of salinity on the first normal stress difference in SAOS flow.
\section{Start-up and cessation of steady shear flow}
\label{Sec:StartupCessation}
\subsection{Analysis}
Start-up and cessation of steady shear flow are two closely related transient shear flows. The velocity field in these flows is described by Eq.  (\ref{Eq:MF:SAOS:VelocityField}); the stress tensor and its Oldroyd derivative are given by Eqs. (\ref{Eq:MF:SAOS:StressTensor}) and (\ref{Eq:MF:SAOS:StressTensorOldroyd}), respectively, while the only independent non-zero component of the rate-of-strain tensor is
\begin{equation}
\label{Eq:MF:SRT:ShearRate}
\dot{\gamma}(t) \equiv \dot{\gamma}_{\,12}(t) = \dot{\gamma}_{\,21}(t) = \dot{\gamma}_{\,0} \Theta(\pm t),
\end{equation} %
\nomenclature[xZi]{$\Theta$}{Heaviside step-function}{-}{first used in Eq. (\ref{Eq:MF:SAOS:ShearRate})}%
where $\dot{\gamma}_{\,0}$ is a constant and $\Theta(t)$ is the Heaviside step-function; for this reason, these flows are known in experimental rheology as step-rate tests. \cite{Mezger2014}
\par 
Choosing the positive sign in Eq. (\ref{Eq:MF:SRT:ShearRate}) corresponds to the start-up case. The fluid is at rest at $t<0$, while a constant shear rate $\dot{\gamma}_{\,0}$ is applied suddenly at $t=0$. After a while, shear and normal stresses build up and approach their steady-shear-flow values.
\par 
In contrast, the negative sign in Eq. (\ref{Eq:MF:SRT:ShearRate}) yields the cessation case, which is the inverse situation. The fluid is flowing steadily with constant shear rate $\dot{\gamma}_{\,0}$ at $t<0$ before the flow is instantaneously stopped (the shear rate is removed) at $t=0$. Shear and normal stresses decay as the fluid approaches equilibrium.
\par 
The material functions of the fluid in this kind of flow -- $\eta^{\pm}$, $\Psi^\pm_1$, and $\Psi^\pm_2$ -- are defined by
\begin{align}
\tau_{12} & = -\eta^\pm(\dot{\gamma}_{\,0},t) \dot{\gamma}_{\,0}, \label{Eq:MF:SRT:DefinitionEtapm} \\
N_1 &= -\Psi^\pm_1(\dot{\gamma}_{\,0},t)\dot{\gamma}_{\,0}^{\,2},\label{Eq:MF:SRT:DefinitionPsi1pm}\\
N_2 &= -\Psi^\pm_2(\dot{\gamma}_{\,0},t)\dot{\gamma}_{\,0}^{\,2} \label{Eq:MF:SRT:DefinitionPsi2pm}
\end{align}%
\nomenclature[sZyhe]{$\eta^-$}{shear stress relaxation function}{Pa\cdot s}{defined by Eq. (\ref{Eq:MF:SRT:DefinitionEtapm})}%
\nomenclature[sZyhf]{$\eta^+$}{shear stress growth function}{Pa \cdot s}{defined by Eq. (\ref{Eq:MF:SRT:DefinitionEtapm})}%
\nomenclature[sZyV6]{$\Psi_1^-$}{first normal stress difference relaxation function}{Pa \cdot s^2}{defined by Eq. (\ref{Eq:MF:SRT:DefinitionPsi1pm})}%
\nomenclature[sZyV7]{$\Psi_1^+$}{first normal stress difference growth function}{Pa \cdot s^2}{defined by Eq. (\ref{Eq:MF:SRT:DefinitionPsi1pm})}%
\nomenclature[sZyV95]{$\Psi_2^-$}{second normal stress difference relaxation \newline function}{Pa \cdot s^2}{defined by Eq. (\ref{Eq:MF:SRT:DefinitionPsi2pm})}%
\nomenclature[sZyV96]{$\Psi_2^+$}{second normal stress difference growth function}{Pa \cdot s^2}{defined by Eq. (\ref{Eq:MF:SRT:DefinitionPsi2pm})}%
and are known as the shear stress, first normal stress, and second normal stress difference growth ($+$) or relaxation ($-$) functions, respectively. When presented graphically, these material functions are commonly normalized to their steady-state values $\eta(\dot{\gamma}_{\,0})$, $\Psi_1(\dot{\gamma}_{\,0})$, and $\Psi_2(\dot{\gamma}_{\,0})$.
\par 
Substituting Eqs. (\ref{Eq:MF:SAOS:StressTensor}), (\ref{Eq:MF:SAOS:StressTensorOldroyd}), and (\ref{Eq:MF:SRT:ShearRate}) into the constitutive equation of the C-FENE-P dumbbells (\ref{Eq:ConstEq:MainConstEq}) and introducing dimensionless quantities
\begin{align}
\mathbb{T}_{ij} &=\dfrac{\tau_{ij}}{nkT}, \label{Eq:MF:SRT:DefinitionTij}\\
r &= t/\lambda, \label{Eq:MF:SRT:Definitionr}\\
\Lambda &= \lambda \dot{\gamma}_{\,0}, \label{Eq:MF:SRT:DefinitionLambda}
\end{align}
where $i,j\in\{1,2,3\}$, lead to a system of nonlinear ordinary differential equations:
\begin{align}
& \dfrac{b}{3}Z \mathbb{T}_{11}+\mathbb{T}'_{11}-2 \mathbb{T}_{12} \Lambda \Theta(\pm r) - (\mathbb{T}_{11}-1)\dfrac{Z'}{Z}=0,  \label{Eq:MF:SRT:DiffEqT11} \\
& \dfrac{b}{3}Z \mathbb{T}_{22}+\mathbb{T}'_{22} - (\mathbb{T}_{22}-1)\dfrac{Z'}{Z}=0,   \label{Eq:MF:SRT:DiffEqT22}\\
& \dfrac{b}{3}Z \mathbb{T}_{33}+\mathbb{T}'_{33} - (\mathbb{T}_{33}-1)\dfrac{Z'}{Z}=0,   \label{Eq:MF:SRT:DiffEqT33}\\
& \dfrac{b}{3}Z \mathbb{T}_{12} + \mathbb{T}'_{12} - \mathbb{T}_{22}\Lambda \Theta(\pm r)-\mathbb{T}_{12} \dfrac{Z'}{Z} =  -\Lambda \Theta(\pm r). \label{Eq:MF:SRT:DiffEqT12}
\end{align}
Here, primes denote differentiation with respect to $r$.
\begin{figure}
\begin{center}
\includegraphics[width=6.45in]{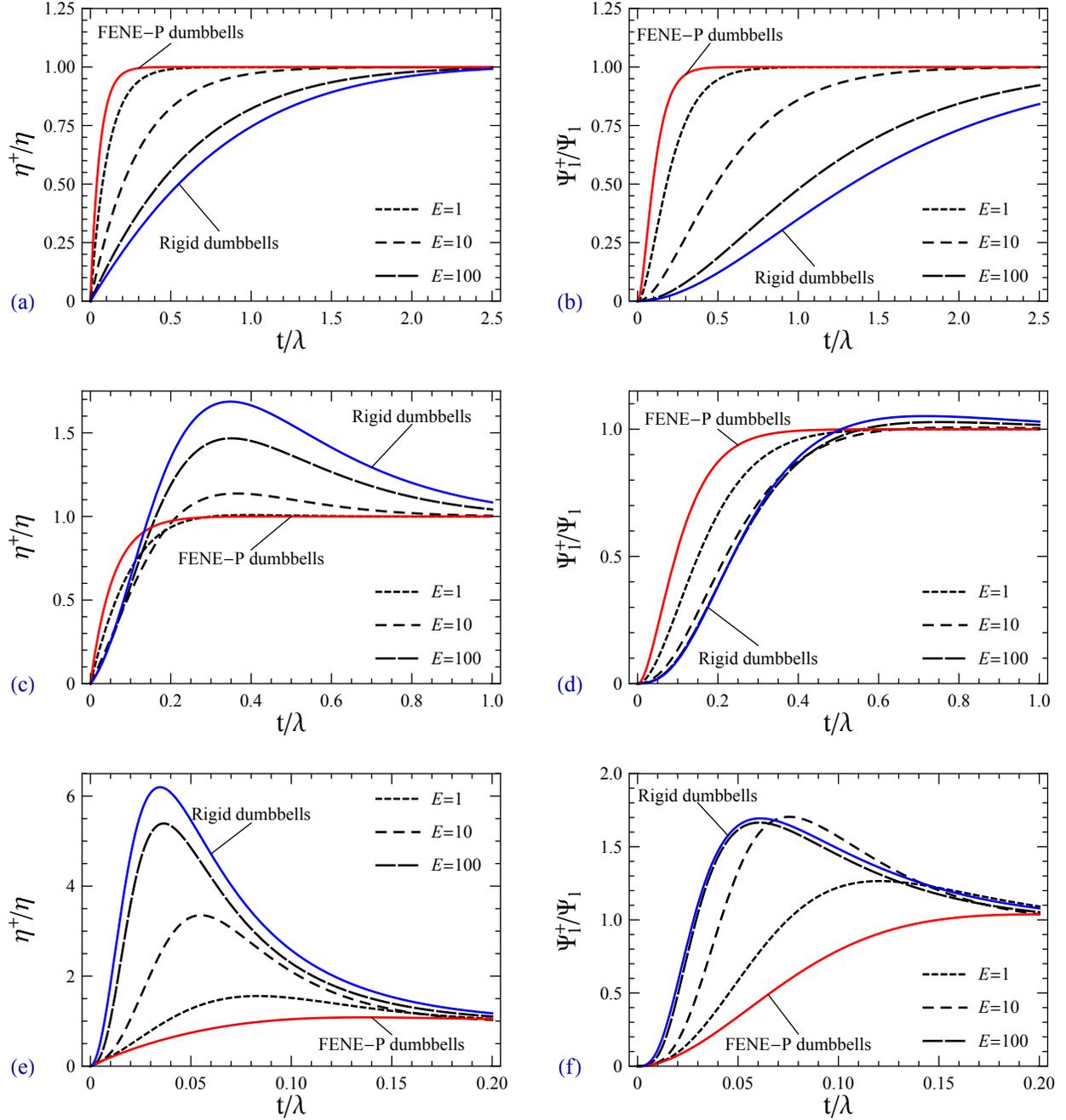}
\caption{Normalized polymer contribution to shear stress [(a), (c), and (e)] and first normal stress difference [(b), (d), and (f)] growth functions of C-FENE-P dumbbells, plotted against dimensionless time, $t/\lambda $, for different values of parameter $E$ at $\lambda \dot{\gamma}_{\,0} = 0.5$ [(a) and (b)], $\lambda \dot{\gamma}_{\,0} = 5$ [(c) and (d)], and $\lambda \dot{\gamma}_{\,0} = 50$ [(e) and (f)]. The limiting cases $E=0$ (FENE-P dumbbells) and $E\to \infty$ (rigid dumbbells) are shown. All curves are plotted at $b=50$. \label{Fig:SRT:Start-up}}
\end{center}
\end{figure}
The expression for the $Z$-factor [Eq. (\ref{Eq:ConstEq:MainZ})] is rewritten accordingly to complete the system. 
The initial conditions are imposed at $r=0$, where the stress tensor components are set to zero (start-up case) or to their steady-shear-flow values (cessation case).
\par 
Since $\mathbb{T}_{22}=\mathbb{T}_{33}=0$ both in equilibrium and in steady shear flow (see Sec. \ref{Sec:SteadyShearFlow}), it follows from Eqs. (\ref{Eq:MF:SRT:DiffEqT22}) and (\ref{Eq:MF:SRT:DiffEqT33}) that $N_2(r)=0$ identically. The rest of the system can be solved numerically. We have used Wolfram Mathematica for this purpose.
\par 
The results for the start-up case are shown in Fig. \ref{Fig:SRT:Start-up}. At very low values of $\Lambda$, the material functions grow seemingly monotonically [see Figs. \ref{Fig:SRT:Start-up}(a) and \ref{Fig:SRT:Start-up}(b)]. At higher $\Lambda$, they typically undergo one or several oscillations around the steady-flow values before they stabilize; a stress overshoot, i.e.,  a time interval where the stresses are higher than their steady-flow values, is clearly seen in Figs. \ref{Fig:SRT:Start-up}(c)-\ref{Fig:SRT:Start-up}(f). Note that the overshoot is not only  observed for rigid dumbbells but also much more pronounced in the RDB limit. Hence, this phenomenon is not caused by dumbbell stretching, as one might have suggested. In addition, it is seen that at higher $\Lambda$, the relative magnitude of overshoots increases and the overshoots are shifted towards earlier times; it takes less time for the stresses to approach their steady-flow values, and for any fixed $\Lambda$, shear stress builds up and stabilizes faster than $N_1$.
\par 
The material functions describing the start-up case depend strongly on $E$. At low $\Lambda$, a decrease in $E$ results in a general increase in $\eta^+/\eta(\dot{\gamma}_{\,0})$ and $\Psi^+_1/\Psi_1(\dot{\gamma}_{\,0})$ [see Figs. \ref{Fig:SRT:Start-up}(a) and \ref{Fig:SRT:Start-up}(b)].  At higher values of $\Lambda$, the impact of $E$ on stress growth becomes more complex. For shear stress growth functions, an overshoot appears [see Figs \ref{Fig:SRT:Start-up}(c) and \ref{Fig:SRT:Start-up}(e)], and as $E$ decreases, the overshoot is shifted towards later times (the higher the $\Lambda$, the more pronounced the effect). At the same time, the relative overshoot magnitude decreases with a decrease in $E$, reaching its minimal value in the FENE-P dumbbell limit ($E=0$). For $N_1$ growth, the situation is somewhat similar [see Figs. \ref{Fig:SRT:Start-up}(d) and \ref{Fig:SRT:Start-up}(f)]; however, the magnitude of $N_1$ overshoot can change non-monotonically with $E$: at high $\Lambda$, it increases, reaches a maximum, and then decreases as $E$ decreases, as shown in Fig. \ref{Fig:SRT:Start-up}(f).
\par 
In other words, an increase in solvent salinity can either increase or decrease $\eta^+/\eta(\dot{\gamma}_{\,0})$, $\Psi^+_1/\Psi_1(\dot{\gamma}_{\,0})$ and the relative $N_1$ overshoot, depending on values of $\Lambda$ and $E$, but always decreases the relative shear stress overshoot.
\begin{figure}
\begin{center}
\includegraphics[width=3.37in]{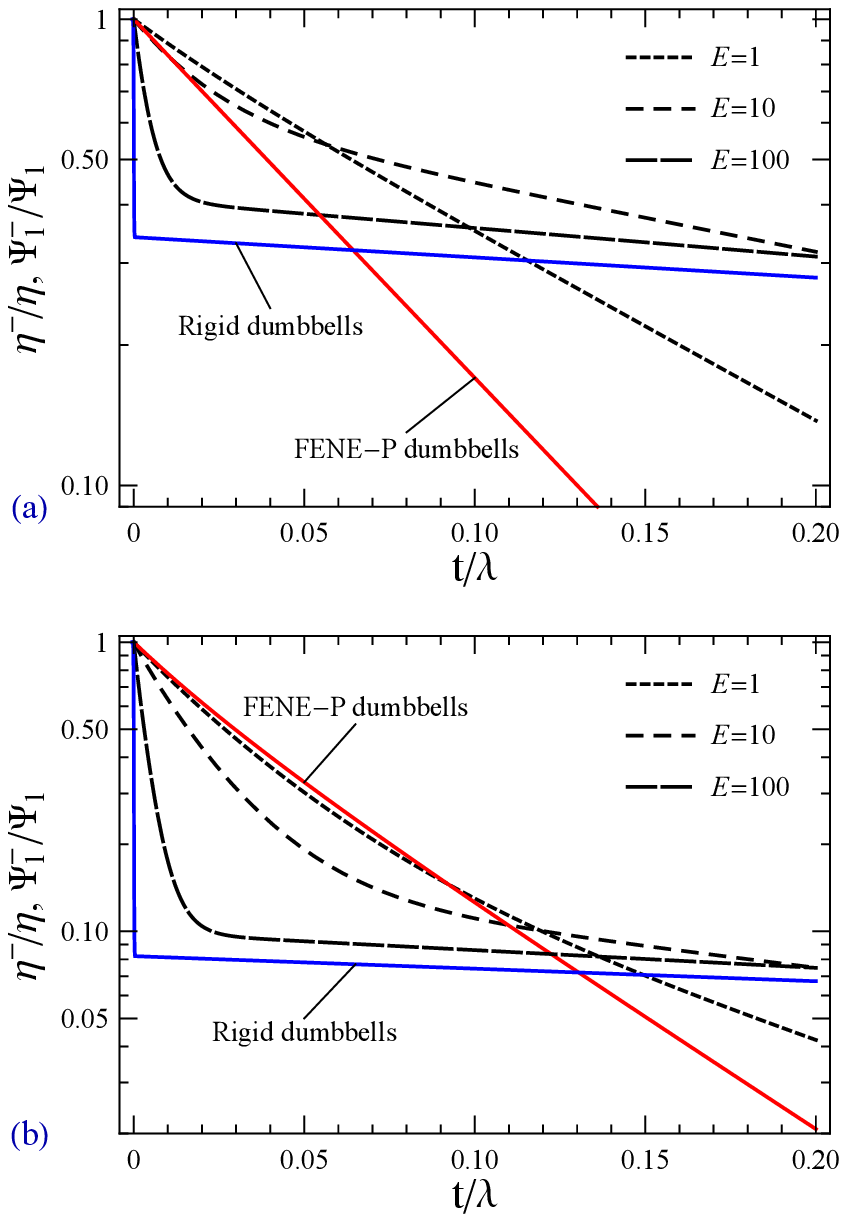}
\caption{Normalized polymer contribution to shear stress and first normal stress difference relaxation functions of C-FENE-P dumbbells, plotted against dimensionless time $ t/\lambda $ at $\lambda \dot{\gamma}_{\,0} = 5$ (a) and $\lambda \dot{\gamma}_{\,0} = 50$ (b) for different values of $E$. The FENE-P limit, $E=0$, and the RDB limit, $E\to \infty$, are also shown. The value of $b$ is set to $50$. \label{Fig:SRT:Relax}}
\end{center}
\end{figure}
\par 
The numerical results for the cessation case are presented in Fig. \ref{Fig:SRT:Relax}. Both the shear stress and the first normal stress difference relaxation functions decay monotonically, quickly approaching zero; the curves for $\eta^-/\eta(\dot{\gamma}_{\,0})$ and $\Psi_1^-/\Psi_1(\dot{\gamma}_{\,0})$ overlap. Equilibrium is approached faster at higher $\Lambda$, as seen from the comparison of Figs. \ref{Fig:SRT:Relax}(b) and \ref{Fig:SRT:Relax}(a).
\par 
At later times, the material functions decay exponentially. In Fig. \ref{Fig:SRT:Relax}, this is seen as regions where the curves become straight lines. The exponential decay is preceded by a region of faster decrease. An analysis of system (\ref{Eq:MF:SRT:DiffEqT11})-(\ref{Eq:MF:SRT:DiffEqT12}) shows that the eigenvalue of its linearized version corresponding to $\mathbb{T}_{12}$ equals $-\mathcal{F}(3/b,E/b)$; hence, 
\begin{equation}
\label{Eq:MF:SRT:Exponent}
\dfrac{\eta^-(t,\dot{\gamma}_{\,0})}{\eta(\dot{\gamma}_{\,0})} \sim \dfrac{\Psi_1^-(t,\dot{\gamma}_{\,0})}{\Psi_1(\dot{\gamma}_{\,0})} \sim \exp[-\mathcal{F}(3/b,E/b)r] \sim \exp (-t/\lambda_\mathrm{\,e})
\end{equation}
asymptotically at later times, where $\lambda_\mathrm{\,e}$ is the time constant defined previously by Eq. (\ref{Eq:MF:SAOS:LambdaE}). Our numerical simulations confirm this analytical result.
\par 
The impact of $E$ on $\eta^-$ and $\Psi_1^-$ is twofold. First, the region of fast decrease at early times, which is abrupt and step-like at large $E$, becomes smoother and less pronounced, as $E$ decreases. Second, a decrease in $E$ leads to a faster decay rate in the exponential regime at late times, since $\mathcal{F}(3/b,E/b)$ is a decreasing function of $E$. Both effects are seen in Fig. \ref{Fig:SRT:Relax}.
\subsection{Discussion}
At fixed salinity, the appearance of a shear stress overshoot at the start-up of steady shear flow and the way its magnitude and position depends on the step-rate value $\dot{\gamma}_{\,0}$ are experimentally observed features of polymer solutions.\cite{Zebrowski1985,Islam2019}
\par 
Our results show that a decrease in salinity leads to a more pronounced shear stress overshoot, an effect that has been  observed experimentally by Zebrowski and Fuller.\cite{Zebrowski1985} 
\par 
In the case of steady shear flow cessation, the C-FENE-P model predicts a faster return to equilibrium at higher shear rates if salinity is held constant, as shown in Fig. \ref{Fig:SRT:Relax}. The same trend is observed in experiments.\cite{Zebrowski1985,Islam2019} Furthermore, our results show that at fixed shear rate value, an increase in salinity leads to a slower decay of shear stresses, which also maches the trend reported by  Zebrowski and Fuller.\cite{Zebrowski1985} 
\par 
Finally, Islam\cite{Islam2019} investigated the shape of normalized shear stress relaxation functions for commercial partially hydrolyzed polyacrylamides. Two  clearly distinct regions -- an exponential decay at later times preceded by a faster decrease at earlier times, as predicted by the C-FENE-P model -- were observed. Furthermore, the decay rate in the exponential regime was found to be dictated by the polymer type and independent of $\dot{\gamma}_{\,0}$ and polymer concentration, which is explained by our theoretical result expressed by Eq. (\ref{Eq:MF:SRT:Exponent}).
\par 
We are not aware of experimental data on the first normal stress difference growth and relaxation functions of polyelectrolyte solutions.
\section{Conclusions}
\label{Sec:Conclusions}
The predictions of the C-FENE-P dumbbell model for steady and transient shear and extensional flows are in a very good qualitative agreement with most of the experimentally observed trends for polyelectrolyte solutions. This involves the shape of the material functions, their dependence on flow parameters, how the material functions are affected by the solvent salinity, and the ways in which the material functions differ for more and less intrinsically rigid polyelectrolytes. At the same time, the C-FENE-P model is strikingly simple, being just slightly more mathematically complex than the original (uncharged) FENE-P dumbbell model. The closed-form constitutive equations allow us to investigate the rheology and fluid dynamics of steady and transient shear and extensional flows using a combination of analytical and simple numerical methods. In addition, the physical quantities of interest, such as the relative extension of the polyelectrolyte molecules, can be easily kept track of under analysis.
\par 
Concerning numerical simulation of complex flows, we assume that computer codes, which are capable of solving the equations of motion for a FENE-P dumbbell fluid, can be relatively easily adapted to tackle the constitutive equations of the C-FENE-P dumbbells. 
\par 
We believe that the C-FENE-P dumbbell model has all the potential to be a robust instrument, suitable for direct use in technological applications and helpful for qualitative understanding of flow phenomena in complex flows of dilute polyelectrolyte solutions.
\begin{acknowledgments}
This research has been funded by VISTA -- a basic research program in collaboration between The Norwegian Academy of Science and Letters and Equinor. The authors acknowledge the Research Council of Norway and the industry partners (2019) of the National IOR Centre of Norway -- ConocoPhillips Skandinavia AS, Aker BP ASA, V{\aa}r Energi AS, Equinor ASA, Neptune Energy Norge AS, Lundin Norway AS, Halliburton AS, Schlumberger Norge AS, Wintershall Norge AS, and DEA Norge AS -- for support. D.S. is thankful to Tamara Shogina for numerous suggestions on improvement of this paper.
\end{acknowledgments}

\section*{Availability of data}
Data sharing is not applicable to this article as no new data were created or analyzed in this study.

\bibliography{NNLib}

\end{document}